\documentclass[12pt]{article}
\usepackage{amsmath}
\usepackage{psfrag,epsf,graphicx}
\usepackage{enumerate}
\usepackage{natbib}
\usepackage{url}

\newcommand{\blind}{0}

\usepackage{multirow,subcaption,amsfonts,stmaryrd,amssymb,amsthm,mathtools,bm,tikz,pifont,XCharter}
\usepackage[ruled]{algorithm2e}
\usetikzlibrary{calc,bayesnet}
\usepackage[backref=page]{hyperref}
\hypersetup{
    colorlinks=true,
    citecolor=blue,
}
\newcommand*{\cond}{\!\mid\!}
\newcommand*{\dquote}[1]{``{#1}"}
\newcommand*{\tr}{\mathsf{T}}
\newcommand*{\indic}[1]{\mathbb{I}{\set{{#1}}}}
\newcommand*{\ind}{\stackrel{ind}{\sim}}
\newcommand*{\iid}{\stackrel{iid}{\sim}}
\newcommand*{\set}[1]{\left\{{#1}\right\}}
\newcommand*{\bracket}[1]{\left({#1}\right)}

\theoremstyle{definition}

\begin{document}

\def\spacingset#1{\renewcommand{\baselinestretch}%
{#1}\small\normalsize} \spacingset{1}

\if0\blind
{
  \title{\bf Adaptive sequential Monte Carlo for structured cross validation in Bayesian hierarchical models}
  \author{
    Geonhee Han\thanks{gh2610@columbia.edu.}\hspace{.2cm} \\
    \small Graduate School of Arts and Sciences, Columbia University\thanks{The majority of this research was carried out while GH was a graduate student at Columbia University GSAS QMSS.} \\
    \small Graduate School of Public Policy, The University of Tokyo \\
    \\
    Andrew Gelman \hspace{.2cm} \\
    \small Department of Statistics and Department of Political Science, Columbia University
    }
    \date{11 Aug 2025}
    \maketitle
} \fi

\if1\blind
{
  \bigskip
  \bigskip
  \bigskip
  \begin{center}
    {\LARGE\bf Adaptive sequential Monte Carlo for cross validation in Bayesian hierarchical models}
\end{center}
  \medskip
} \fi

\bigskip

\begin{abstract}

Importance sampling (IS) is commonly used for cross validation (CV) in Bayesian models, because it only involves reweighting existing posterior draws without needing to re-estimate the model by re-running Markov chain Monte Carlo (MCMC).
For hierarchical models, standard IS can be unreliable;
the out-of-sample generalization hypothesis may involve structured case-deletion schemes which significantly alter the posterior geometry.
This can force costly MCMC re-runs and make CV impractical.
As a principled alternative, we tailor adaptive sequential Monte Carlo to sample along a path of posteriors that leads to the case-deleted posterior.
The sampler is designed
to support various hypotheses by accommodating diverse CV designs, and
to streamline the workflow by automating path construction and systematically minimizing MCMC intervention.
We demonstrate its utility with three types of predictive model assessment:
longitudinal leave-group-out CV,
group $K$-fold CV, and
sequential one-step-ahead validation.

\end{abstract}

\noindent%
\vfill

\newpage
\spacingset{1.6}
\renewcommand{\arraystretch}{0.6}


\section{Introduction}

Evaluating the fit of a Bayesian model by identifying discrepancies between the model and the data is a crucial step of the Bayesian workflow \citep{GelmanVehtariSimpsonMargossianCarpenterYaoKennedyGabryBurknerModrak2020}.
In particular, \textit{predictive model assessment} focuses on how well a model can predict new and unseen data, often assessed via cross validation \citep{Stone1976, Geisser1975, GeisserEddy1979, ArlotCelisse2010, VehtariOjanen2012, PiironenVehtari2016}.

With Bayesian models, cross validation (CV) is known to be computationally intensive due to the need for re-estimating the posterior distributions for datasets that omit subsets of observations.
For example, na\"ive leave-one-out cross validation (LOO-CV) requires separate posterior estimations for each omitted observation, typically performed using computationally expensive methods such as Markov chain Monte Carlo (MCMC);
this makes na\"ive LOO-CV computationally impractical for large datasets or complex models.
{A popular remedy is to use importance sampling (IS) and its variants} \citep{GelfandDey1994, Peruggia1997, EpifaniMacEachernPeruggia2008, VehtariGelmanGabry2017, LoboFonsecaMoura2020}, which approximate the case-deleted posterior by re-weighting posterior samples obtained from the full dataset, circumventing the need for repeated re-estimation and providing substantial computational savings.

There is considerable interest in efficient CV for structured Bayesian hierarchical models, such as those with longitudinal, spatial, or temporal structure.
For example, with models for grouped data, identifying groups that are challenging to predict with leave-group-out CV \citep{MerkleFurrRabe-Hesketh2019, LiuRue2023, AdinTeixeiraLenziLiuMartínez-MinayaRue2024, ZhangDanielsLiBao2024} can highlight specific groups where the hierarchical model struggles to predict and motivate model expansions \citep[Chap.~6.2]{GelmanVehtariSimpsonMargossianCarpenterYaoKennedyGabryBurknerModrak2020}.

A potential challenge with IS for such Bayesian models is its instability in estimates, such as due to possibly infinite variance in importance weights \citep{VehtariGelmanGabry2017, Millar2018, SilvaZanella2023, ChangLiXuYaoPorcinoChow2024}.
{CV in structural Bayesian models often requires non-standard design of blocking structures and out-of-sample prediction schemes that account for intricate dependencies} \citep{GelmanHwangVehtari2014, RobertsBahnCiutiBoyceElithGuillera-ArroitaHauensteinLahoz-MonfortSchröderThuillerWartonWintleHartigDormann2017}.
Such case-deletion schemes can result in distant posteriors that
(a) a vanilla IS estimator would struggle to approximate accurately and reliably, and
(b) would inevitably necessitate additional runs of MCMC to re-approximate the case-deleted posterior, which is extremely impractical.
Some examples are
spatial, temporal, and nested multilevel structures (e.g., phylo-genetic models) which involve dependent observations that are highly informative to the posterior geometry:
see \cite{BurknerGabryVehtari2020}, \cite{BürknerGabryVehtari2021}, \cite{LoboFonsecaMoura2020}, and \cite{Martínez-MinayaHaavard2024}.

Research on computational methods to efficiently perform CV with structural blocking or case-deletion schemes remains limited.
Recent work by \cite{LiuRue2023} introduced methods for approximating a leave-group-out estimand in {latent Gaussian models, leveraging the conditional independence of observations given linear Gaussian predictors.
Their approach uses direct numerical integration by exploiting the inherent tractability of latent Gaussian models.}
In Bayesian hierarchical models, \cite{ZhangDanielsLiBao2024} focus on estimating cross validated means rather than the (log) predictive density.
Mixture estimators have been introduced by \cite{SilvaZanella2023} for the computation of LOO-CV estimands, where the asymptotic variance of the weights is finite, although additional simulation from a proposal (often of a non-standard form) is required, essentially necessitating re-runs of, say, MCMC.
Efforts to avoid MCMC re-runs through moment matching were explored by \cite{PaananenPiironenBürknerVehtari2021, VehtariBürknerGabry2024}, while the authors also concede that affine transformation may be insufficient to produce suitable proposals and suggest that more complex methods may be needed, beyond LOO-CV.
Other existing work explores case-deleted posterior approximations using a local sensitivity approach for sensitivity analysis \citep{GhoshStephensonNgyuenDeshpandeBroderick2020, BroderickGiordanoMeager2023, NguyenGiordanoMeagerBroderick2024, HuangBurtNgyuenShenbroderick2024}, rather than model evaluation.

Our goal is to develop a computational approach applicable to a wide range of structural Bayesian models and CV schemes, which can be executed as a byproduct of a single MCMC run on a full non-case-deleted dataset, complementing the prominent MCMC-based Bayesian workflow.
The method adopts the adaptive sequential Monte Carlo (aSMC) sampler \citep{DelMoralDoucetJasra2006, JasraStephensDoucetTsagaris2011}, and bridges distant posteriors by automatically constructing a sequence of auxiliary intermediate distributions leading to the target case-deleted posterior(s).
The sampler is applicable to a wide range of models and CV schemes, while allowing one to avoid the costly MCMC re-runs whenever possible, and further being equipped with design-efficient sample-generating capabilities, unlike existing IS methods, even when the target posteriors are detected to be distant.

The structure of the paper is as follows.
Section \ref{sec:2} describes the setup and explores various structural CV schemes.
In Section \ref{sec:3}, we consider the aSMC approach.
Section \ref{sec:4} demonstrates the application of the method in three examples involving grouped, time-series, and spatial data.
Section \ref{sec:5} concludes with remarks and discussions.

\section{Predictive evaluation of Bayesian hierarchical models} \label{sec:2}

\subsection{Bayesian hierarchical model}

We index the groups by $g \in \set{1, \ldots, G}$ and
the observations in each group by $i \in \set{1, \dots N_g}$.
Consider a Bayesian hierarchical model where $y_{g,i}$ represents the $i$-th observation within group $g$ \citep[\'{a} la][Section 5.2]{GelmanCarlinSternDunsonVehtariRubin2020}, defined as
\begin{equation*}
    \phi \ind p(\cdot),
    \qquad
    \theta_g \cond \phi \ind p(\cdot \cond \phi),
    \qquad
    y_{g,i} \cond \theta_g, \phi \ind p(\cdot \cond \theta_g, \phi),
\end{equation*}
where $\phi$ is a global parameter (hyperprior) and $\theta_g$ is a group-specific parameter.
None are restricted to being univariate.
The posterior distribution of the parameters $\bm{\Theta} = (\phi, \theta_{1:G})$ is proportional to the joint distribution
\begin{equation*}
    p(\phi)
    \prod_{g=1}^G p(\theta_g \cond \phi)
    \prod_{i=1}^{N_g} p(y_{g,i} \cond \theta_g, \phi),
\end{equation*}
up to a normalizing constant.
We assume that $y \mapsto p(y_{g,i} = y \cond \theta_g, \phi)$ may be evaluated.

\paragraph{Examples.}

Some examples of models which may involve non-standard structural CV schemes are as follows.
\begin{itemize}
    \item \textbf{Grouped models}:
    With grouped or panel data, group-specific parameters capture variation across units in a group or over time,
    \begin{equation*}
        y_{g,i} = \bm{x}_{g,i}^\tr \bm{\beta}_g + \varepsilon_{g,i}
        ,
    \end{equation*}
    where $\bm{\beta}_g \ind p(\cdot)$ represents the coefficients specific to group $g$, and $\bm{x}_{g,i}$ denotes covariates.
    $\varepsilon_{g,i}$ are conditionally independent,
    and the conditional likelihood of $y_{g,i}$ is evaluable.

    \item \textbf{(Hierarchical) spatial regression}:
    Structured covariation within groups, such as spatial correlation, may involve
    \begin{equation*}
        \bm{y}_{g,i} = \bm{X}_{g,i} \bm{\beta}_g + \bm{\omega}_{g,i} + \bm{\varepsilon}_{g,i},
    \end{equation*}
    Spatial dependence may involve, e.g.,
    \begin{equation*}
        \bm{\omega}_{g,i} \ind \mbox{MVN}(\bm{0}, \bm{V}_g), \qquad \bm{\varepsilon}_{g,i}
        \ind \mbox{MVN}(\bm{0}, \sigma^2 \bm{I}),
    \end{equation*}
    where
    {$\mbox{MVN}(\bm{\mu}, \bm{\Sigma})$ is the multivariate normal distribution with mean and covariance $(\bm{\mu}, \bm{\Sigma})$},
    and covariance $\bm{V}_g$ expresses the $g$-specific intra-dependency.
    Here, we treat the $(g,i)$-th response as multivariate.
    
    \item \textbf{Dynamic normal linear models}:
    Temporal dynamics within groups may involve
    \begin{align*}
        \bm{y}_{g,t} &= \bm{X}_{g,t} \bm{\beta}_{g,t} + \bm{\varepsilon}_{g,t}^{(y)}, \qquad
        \bm{\varepsilon}_{g,t}^{(y)} \ind \mbox{MVN}(\bm{0}, \bm{\Sigma}), \\
        \bm{\beta}_{g,t} &= \bm{\beta}_{g,t-1} + \bm{\varepsilon}_{g,t}^{(\beta)}, \qquad
        \bm{\varepsilon}_{g,t}^{(\beta)} \ind \mbox{MVN}(\bm{0}, \bm{V}),
    \end{align*}
    and $\bm{\beta}_{g,0} \ind p(\cdot)$.
    Here, we have identified
    $\phi := (\bm{\Sigma}, \bm{V})$ and
    $\theta_g := \bm{\beta}_{g,0:T} := (\bm{\beta}_{g,0}, \ldots, \bm{\beta}_{g,T})^\tr$, and $\bm{\beta}_{g,t}$ evolves smoothly over time with $p(\theta_g \cond \phi) = p(\bm{\beta}_{g,0} \cond \phi) \prod_{t=1}^T p(\bm{\beta}_{g,t} \cond \bm{\beta}_{g,t-1}, \bm{V})$.
\end{itemize}

\subsection{Case-deletion schemes and computation} \label{sec:2:2}

A widely used method to evaluate the fit of a Bayesian model is to assess its out-of-sample predictive performance \citep{Roberts1965, Guttman1967, GeisserEddy1979, VehtariOjanen2012}.
One of many approaches is within-sample CV \citep{Stone1977}, with advancements in computationally efficient techniques {such as approximate LOO-CV} using IS \citep{GelfandDey1994, Peruggia1997, EpifaniMacEachernPeruggia2008, VehtariGelmanGabry2017}.
For a more detailed overview of these methods, see \cite{VehtariMononenTolvanenSivulaWinther2016}.
We present examples of possible structural schemes in Bayesian hierarchical models below to provide an overview and highlight potential computational challenges associated with structural CV.

\subsubsection{Leave one-in-group out (LOO)} \label{sec:2:2:1}

The LOO-CV scheme, as outlined in \cite{VehtariGelmanGabry2017}, can be applied to the above Bayesian hierarchical model as follows.
The leave-$(h,j)$-out posterior, corresponding to excluding the observation $y_{h,j}$ with $h \in \set{1, \ldots, G}$ and $j \in \set{1, \ldots, N_h}$, is defined as:
\begin{equation*}
    p_{-(h,j)}(y^*, \bm{\Theta})
    \propto p(\phi)
    \prod_{g=1}^G p(\theta_g \cond \phi)
    \prod_{i=1}^{N_g}
    p(y_{g,i} \cond \theta_g, \phi)^{\indic{(g,i) \neq (h,j)}}
    p(y^* \cond \theta_g, \phi)^{\indic{(g,i) = (h,j)}},
\end{equation*}
where $y^*$ is the posterior predictive for observation $(h,j)$, treated as unobserved, along with the model parameters $\bm{\Theta}$.

Under the logarithmic scoring rule \citep{GneitingRaftery2007}, out-of-sample predictive accuracy for the excluded unit is evaluated via its log posterior predictive distribution.
Under the hierarchical model, the posterior predictive distribution for a new replication within group $g$ under the leave-$(g,i)$-out posterior is obtained by integrating out the parameters $\bm{\Theta}$,
\begin{equation*}
    p_{-(g,i)}(y)
    = \int p_{-(g,i)}(y_{g,i}^* = y, \bm{\Theta}) \; d\bm{\Theta}.
\end{equation*}
Then, given the collection $(y_{g,i})_{g,i}$, the log pointwise predictive density for \textit{new} within-group observations (\citealp{VehtariGelmanGabry2017}; \citealp[Chap.~7]{GelmanCarlinSternDunsonVehtariRubin2020}) would be
\begin{equation*}
    \ell^{(\text{LOO})}
    := \sum_{g=1}^G \sum_{i=1}^{N_g} \log p_{-(g,i)}(y_{g,i}).
\end{equation*}

A na\"ive approach to compute the estimand requires (re-)executing posterior inference $N_1 + \ldots + N_G$ times, once for each leave-$(g,i)$-out posterior.
This strategy is clearly computationally expensive and impractical.
A more efficient strategy uses importance weighting, taking advantage of posterior samples from the \textit{baseline} non-case-deleted posterior.
One in practice approximates the leave-$(g,i)$-out posterior by giving the draws $\bm{\Theta}^{(r)} = (\phi^{(r)}, \theta_{1:G}^{(r)}) \sim p(\cdot \cond \bm{y}_{1:G})$ over $r = 1, \ldots, R$ ({where $R$ is the number of posterior MCMC draws}) an importance ratio
\begin{equation*}
    w_{g,i}^{(r)} := p(y_{g,i} \cond \theta_g^{(r)}, \phi^{(r)})^{-1},
    \qquad
    W_{g,i}^{(r)} := {w_{g,i}^{(r)} \over \sum_{r=1}^R w_{g,i}^{(r)}},
\end{equation*}
where $w_{g,i}^{(r)}$ is the $(g,i)$-importance ratio and $W_{g,i}^{(r)}$ its self-normalized weight for the $r$-th draw.
The LOO estimand is then approximated by the right-hand side approximation,
\begin{equation}
    \ell^{(\text{LOO})}
    \approx \sum_{g=1}^G \sum_{i=1}^{N_g} \log \sum_{r=1}^R W_{g,i}^{(r)} p(y_{g,i}^* = y_{g,i} \cond \theta_g^{(r)}, \phi^{(r)})
    =: \hat{\ell}^{(\text{LOO})}, \label{eq:loo-lpd-est}
\end{equation}
which is computable if the predictive distribution in the summand over $(g,i)$ can be evaluated, which is usually true.

\subsubsection{Leave group out (LGO)} \label{sec:2:2:2}

Case deletion in hierarchical models can extend beyond individual within-group observations to entire groups (e.g., Section \ref{sec:4:radon}).
The leave-$g$-out posterior, where $g$ denotes the excluded group, is defined as
\begin{equation*}
    p_{-(g,:)}(y_{1:N_g}^*, \bm{\Theta})
    \propto
    p(\phi)
    \prod_{h \neq g} p(\theta_{h} \cond \phi)
    \prod_{i=1}^{N_g}
        p(y_{h,i} \cond \theta_{h}, \phi)^{\indic{h \neq g}}
        p(y_i^* \cond \theta_{h}, \phi)^{\indic{h = g}}.
\end{equation*}
In other words, we condition on all but $\bm{y}_g = y_{g,1:N_g} = (y_{g,1}, \ldots y_{g,N_g})^\tr$, which we simply write as $\bm{y}_{-g}$.
Using this distribution, the posterior predictive density $p_{-(g,:)}(\cdot)$, evaluated at $\bm{y}_g$, defines a new estimand,
\begin{equation}
    \ell^{(\text{LGO})}
    := \sum_{g=1}^G \log p_{-(g,:)}(\bm{y}_g),
    \label{eq:lgo-lpd-est}
\end{equation}
where $p_{-(g,:)}(\cdot)$ is the joint posterior predictive distribution marginalized over the leave-$g$-out posterior,
\begin{equation*}
    p_{-(g,:)}(\bm{y}_g)
    = \int p(\bm{y}_g^* = \bm{y}_g \cond \theta_g, \phi) p(\bm{\Theta} \cond \bm{y}_{-g}) \; d\bm{\Theta}.
\end{equation*}
The procedure is also described by \cite{MerkleFurrRabe-Hesketh2019} as the \textit{approximate leave-one-cluster-out CV}.
Unlike the LOO scheme for hierarchical models, which evaluates individual conditionally independent observations within a group (Section \ref{sec:2:2:1}), LGO log pointwise predictive density assesses the joint predictive accuracy for a hypothetical replication of the entire group.

Under this setup, the unnormalized importance ratio is
\begin{equation*}
    w_g
    := p(\bm{y}_g \cond \theta_g, \phi)^{-1}
    = \prod_{i=1}^{N_g} p(y_{g,i} \cond \theta_g, \phi)^{-1},
\end{equation*}
where the second equality follows from the conditional independence of within-group observations $\bm{y}_g$.
An estimate of the LGO log pointwise predictive density is computed by weighting then aggregating the respective joint predictive density,
\begin{equation}
    \ell^{(\text{LGO})}
    \approx \hat{\ell}^{(\text{LGO})}
    := \sum_{g=1}^G \hat{\ell}_g^{(\text{LGO})}
    := \sum_{g=1}^G \log \sum_{r=1}^R W_g^{(r)} p(\bm{y}_g^* = \bm{y}_g \cond \theta_g^{(r)}, \phi^{(r)})
    .
    \label{eq:lgo-lpd-est:is}
\end{equation}
The self-normalized weights $W^{(r)}$ are computed as before.

\subsubsection{Backward-sequential leave end out (LEO)} \label{sec:2:2:3}

For settings with inherent ordering (e.g., time-series data under dynamic models, in Section \ref{sec:4:yield}), it often makes sense to align blocking schemes with that ordering to better assess predictive performance on future sequentially arriving data.
Below are non-exhaustive examples of backward-sequential exclusion of the most recent observations.

\paragraph{Within-group.}

Fix a group index $g$, and assume that $i$ indexes time.
We define the sequential LEO posterior for group $g$,
which conditions upon all of $\bm{y}_{-g}$ and the subset $y_{g,1:t} = (y_{g,1}, \ldots, y_{g,t})^\tr$ up to time $t$,
\begin{equation*}
    p_{-(g,~t+1:T_g)}(y_{t+1:T_g}^*, \bm{\Theta})
    \propto p(\phi)
    \prod_{h = 1}^G p(\theta_{h} \cond \phi)
    \prod_{i=1}^{T_{h}}
        \begin{cases}
            p(y_{i}^* \cond \theta_{h}, \phi) & \mbox{if } (h = g) \mbox{ and } (i \in (t+1:T_h)))\\
            p(y_{h,i} \cond \theta_h, \phi) & \text{otherwise}
        \end{cases}.
\end{equation*}
$T_g = N_g$ now represents the horizon.
We then reverse the time index as $t = T_g-1, \ldots, 0$ for backward-sequential exclusion.

The unnormalized importance weight associated with this LEO scheme given $t$ is $w_g = \prod_{i=t+1}^{T_g} p(y_{g,i} \cond \theta_g, \phi)^{-1}$.
A natural estimand is the multi-step ahead log pointwise predictive density,
\begin{align*}
    \ell_g^{(\text{LEO})}
    &:= \log
        \underset{\bm{\Theta}, y^*}{\mathbb{E}}\!\!
            \left(
                p_{-(g,~t+1:T_g)}(
                    y_{t+1}^*, \;
                    \ldots, \;
                    y_{t+h-1}^*, \;
                    \underbrace{y_{t+h}^* = y_{g,t+h}}_{\text{evaluate }y_{g,t+h}}, \;
                    \bm{\Theta}
                )
                \;\Big|\;
                \bm{y}_{-g}, y_{g,1:t}
            \right) \\
    &\approx
        \log
        \sum_{r=1}^R W_g^{(r)}
        p_{-(g,~t+1:t+h)}(
            [y_{(t+1):(t+h-1)}^*]^{(r)}, \;
            y_{t+h}^* = y_{g,t+h}, \; \bm{\Theta}^{(r)}
        )
    =: \hat{\ell}_g^{(\text{LEO})},
\end{align*}
after marginalizing out the posterior predictive after $y_{t+h}^*$.
The operator $\underset{\bm{\Theta}}{\mathbb{E}}$ denotes expectation with respect to $\bm{\Theta}$.
In most modeling situations, it would likely be the case that $G=1$, and a joint model (\citealp[e.g., dynamic generalized linear model:][]{WestHarrisonMigon1985}; \citealp[Chap.~16]{WestHarrison1997}) specifies the inter-temporal and inter-variable dependence of the observed multivariate sequence.

\paragraph{Across-group.}

For $G > 1$, and assuming that all groups have equal trajectory lengths $T_g = T$ for simplicity, the LEO posterior can be generalized across all groups by similarly indexing backward as $t = T-1, \ldots, 0$ and defining the posterior by leaving out $(y_{1,t+1:T}, \ldots, y_{G,t+1:T})$, where the importance weight for this scheme is the product $\prod_{g=1}^G w_g = \prod_{g=1}^G \prod_{i=t}^{T} p(y_{g,i} \cond \theta_g, \phi)^{-1}$ which leads to the across-group evaluation $\ell^{(\text{LEO})} := \sum_{g=1}^G \ell_g^{(\text{LEO})}$.

The above are illustrative cases;
more general schemes may be relevant when group-level series vary in length and/or are sampled at mixed temporal resolutions/frequencies, and these should be explicitly accommodated in the case-deletion scheme beyond the standard scheme.

\subsubsection{Leave subset out (LSO)} \label{sec:2:2:4}

The preceding CV schemes can be generalized by defining a set of indices $\mathcal{I}_k \subseteq \mathcal{I} := \bigcup_{g = 1}^G \set{g} \times \set{1, \ldots, N_g}$ {at which the corresponding observations are deleted} from the baseline posterior.
The unnormalized importance weight for this general case is $w_k := \prod_{(g,i) \in \mathcal{I}_k} p(y_{g,i} \mid \theta_g, \phi)^{-1}$.

Some further specific examples are \textit{$K$-fold} or \textit{group $K$-fold CV}, where the observations $(y_{g,i})_{g,i}$ are divided into $K$ mutually exclusive partitions such that $\mathcal{I} = \bigsqcup_{k=1}^K \mathcal{I}_k$.
The group $K$-fold is implemented by considering partitions $\mathcal{I}_{1:K}$ which appropriately accounts for strata or grouping structure (e.g., group $K$-fold in Section \ref{sec:4:m5}).

Multiple groups $\varnothing \subsetneq \mathcal{G} \subsetneq \set{1, \ldots, G}$ may likewise be excluded in what we can term \textit{leave-groups-out}, which induces posterior predictive distributions and corresponding replications over multiple groups.

\paragraph{Agenda}

We illustrated that non-standard case-deletion schemes can yield importance weights formed by reciprocals of products of conditional likelihoods, and that this arises naturally in Bayesian hierarchical models and their (C)V schemes, such as LGO, LEO, and LSO (including $K$-fold and group $K$-fold).
Instability is expected in these settings, especially when, as is likely, the baseline posterior exhibits thinner tails than its case-deleted counterparts\citep[e.g., leading to high- or infinite-variance weights:][]{EpifaniMacEachernPeruggia2008, VehtariGelmanGabry2017, SilvaZanella2023, VehtariSimpsonGelmanYaoGabry2024}.
In the next section, we design an {aSMC sampler} that automatically constructs a sequence of intermediate distributions bridging the baseline and the target distribution, to facilitate a stable approximation of geometries that are otherwise difficult to traverse directly.

\section{Sequential Monte Carlo approach} \label{sec:3}

\subsection{Bridging distant posteriors via Markov kernels} \label{sec:3:1}

Let the baseline unnormalized posterior be
\begin{equation*}
    \gamma_0(\bm{\Theta})
    := p(\phi)
    \prod_{g=1}^G p(\theta_g \cond \phi)
    \prod_{i=1}^{N_g} p(y_{g,i} \cond \theta_g, \phi),
\end{equation*}
and the target unnormalized posterior be
\begin{equation*}
    \gamma_{k}(\bm{\Theta})
    := p(\phi)
    \prod_{g=1}^G p(\theta_g \cond \phi)
    \prod_{i=1}^{N_g} p(y_{g,i} \cond \theta_g, \phi)^{\indic{(g,i) \notin \mathcal{I}_k}}.
\end{equation*}
The index $k \in \set{1, \ldots, K}$ references the set of observation indices $\mathcal{I}_k$ that are to be deleted from the baseline posterior (e.g., LGO with $\mathcal{I}_g = \set{g} \times \set{1, \ldots, N_g}$).
Then, induce $K$ targets $p_1, \ldots, p_K$ from the respective unnormalized posteriors,
\begin{equation*}
    p_k(\bm{\Theta}) = {\gamma_k(\bm{\Theta}) \over Z_k},
\end{equation*}
where $Z_k = \int \gamma_k(\bm{\Theta}) \; d\bm{\Theta}$ is an unknown normalizing constant.

For each of these $K$ targets, we now prepare a sequence of (intermediate) distributions $\gamma_{k,\ell}$ for $\ell \in \set{0, 1, \ldots, L_k}$  such that $\gamma_{k,0} = \gamma_0$ and $\gamma_{k,L_k} = \gamma_k$.
Fixing $k$ henceforth and following \cite{DelMoralDoucetJasra2006}, in a common product space, we introduce backward Markov kernels $(\mathcal{L}_{k,\ell-1})_{\ell=1}^{L_k}$ as
\begin{equation*}
    \tilde{p}_k(\bm{\Theta}_0, \ldots, \bm{\Theta}_{L_k})
    := p_{k,L_k}(\bm{\Theta}_{L_k}) \prod_{\ell=1}^{L_k} \mathcal{L}_{k,\ell-1}(\bm{\Theta}_{\ell-1} \leftarrow \bm{\Theta}_{\ell}),
\end{equation*}
which admits $p_{k,L_k} = p_{k}$ as its marginal with respect to $\bm{\Theta}_{L_k}$.
We then introduce forward Markov kernels $(\mathcal{K}_{k,\ell})_{\ell=1}^{L_k}$ such that
\begin{equation*}
    \tilde{q}_k(\bm{\Theta}_0, \ldots, \bm{\Theta}_{L_k})
    := p_{k,0}(\bm{\Theta}_0) \prod_{\ell=1}^{L_k} \mathcal{K}_{k,\ell}(\bm{\Theta}_{\ell} \leftarrow \bm{\Theta}_{\ell-1}).
\end{equation*}
It follows from the Radon--Nikodym theorem that
\begin{equation}
    \underset{\tilde{p}_k}{\mathbb{E}}(f(\bm{\Theta}_{\ell}))
    = \underset{\tilde{q}_k}{\mathbb{E}}\!\left(
        f(\bm{\Theta}_{\ell})
        {\tilde{p}_k \over \tilde{q}_k}(\bm{\Theta}_0, \ldots, \bm{\Theta}_{\ell})
    \right),
    \label{eq:smc:radon-nikodym}
\end{equation}
where ${\tilde{p}_k / \tilde{q}_k}$ is as follows.
First, define the forward kernels $(\mathcal{K}_{k,\ell})_{\ell=1}^{L_k}$ as invariant kernels (e.g., MCMC) targeting the respective intermediate distributions $(p_{k,\ell})_{\ell=1}^{L_k}$.
Further define the backward kernels as the time reversal of forward kernels, that is $\mathcal{L}_{k,\ell-1}(\bm{\Theta}_{\ell-1} \leftarrow \bm{\Theta}_{\ell}) := \mathcal{K}_{k,\ell}(\bm{\Theta}_{\ell} \leftarrow \bm{\Theta}_{\ell-1}) {\gamma_{k,\ell}(\bm{\Theta}_{\ell-1}) / \gamma_{k,\ell}(\bm{\Theta}_{\ell})}$, we obtain incremental weights \citep{DaiHengJacobWhiteley2022} within the expectation of the form
\begin{equation*}
    {\tilde{p}_k \over \tilde{q}_k}(\bm{\Theta}_0, \ldots, \bm{\Theta}_{\ell})
    \propto
    \prod_{l=1}^{\ell} {\gamma_{k,l}(\bm{\Theta}_{l}) \over \gamma_{k,l-1}(\bm{\Theta}_{l-1})} {\mathcal{L}_{k,l-1}(\bm{\Theta}_{l-1} \leftarrow \bm{\Theta}_{l}) \over \mathcal{K}_{k,l}(\bm{\Theta}_{l} \leftarrow \bm{\Theta}_{l-1})}
    =: \prod_{l=1}^{\ell} w_{k,l}(\bm{\Theta}_{l-1}),
\end{equation*}
where we have defined
\begin{equation}
    w_{k,\ell}(\bm{\Theta}_{\ell-1})
    := {\gamma_{k,\ell}(\bm{\Theta}_{\ell-1}) \over \gamma_{k,\ell-1}(\bm{\Theta}_{\ell-1})}.
    \label{eq:smc:weight}
\end{equation}
We therefore arrive at an approximately executable sampler over the augmented space $\bm{\Theta}_{0:L_k}$ provided access to draws from the baseline distribution $p_{k,0}$.
The access holds in practice, and aligns with the usual computational setup, in the sense that a single \textit{baseline} draw from the complete non-case-deleted posterior is available from established tools such as \texttt{Stan} \citep{CarpenterGelmanHoffmanLeeGoodrichBetancourtBrubakerGuoLiRiddell2017}.
Note that we are abbreviating the dependence of the parameters on the index $k$ for simplicity because this is fixed.

\subsection{Parameterizing case deletions} \label{sec:3:2}

\begin{figure}
    \centerline{
    \subfloat[Leave-group-out (LGO)]{\includegraphics[width=0.33\textwidth]{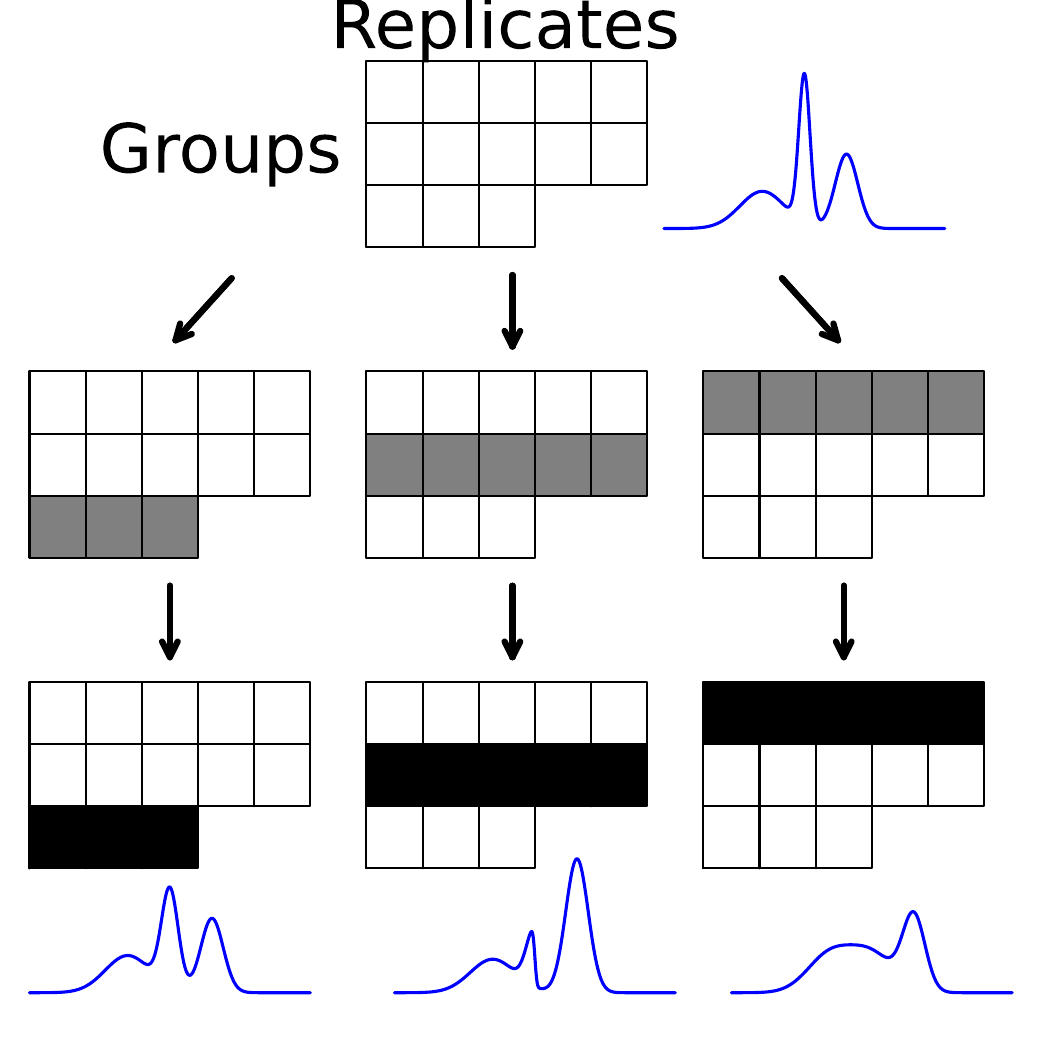} \label{fig:smc-example:lgo}} ~
    \subfloat[Leave-subset-out (LSO)]{\includegraphics[width=0.33\textwidth]{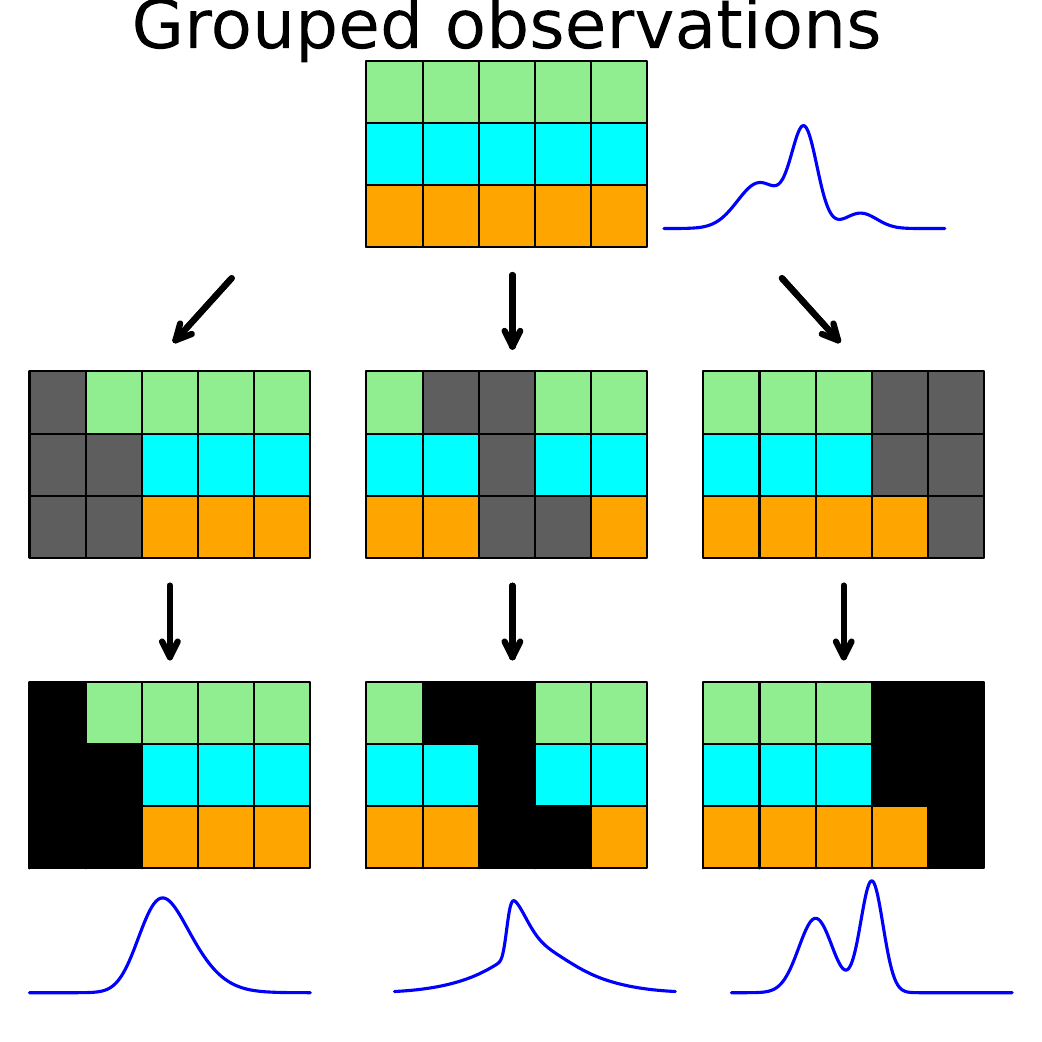} \label{fig:smc-example:lso}} ~
    \subfloat[Leave-end-out (LEO)]{\includegraphics[width=0.33\textwidth]{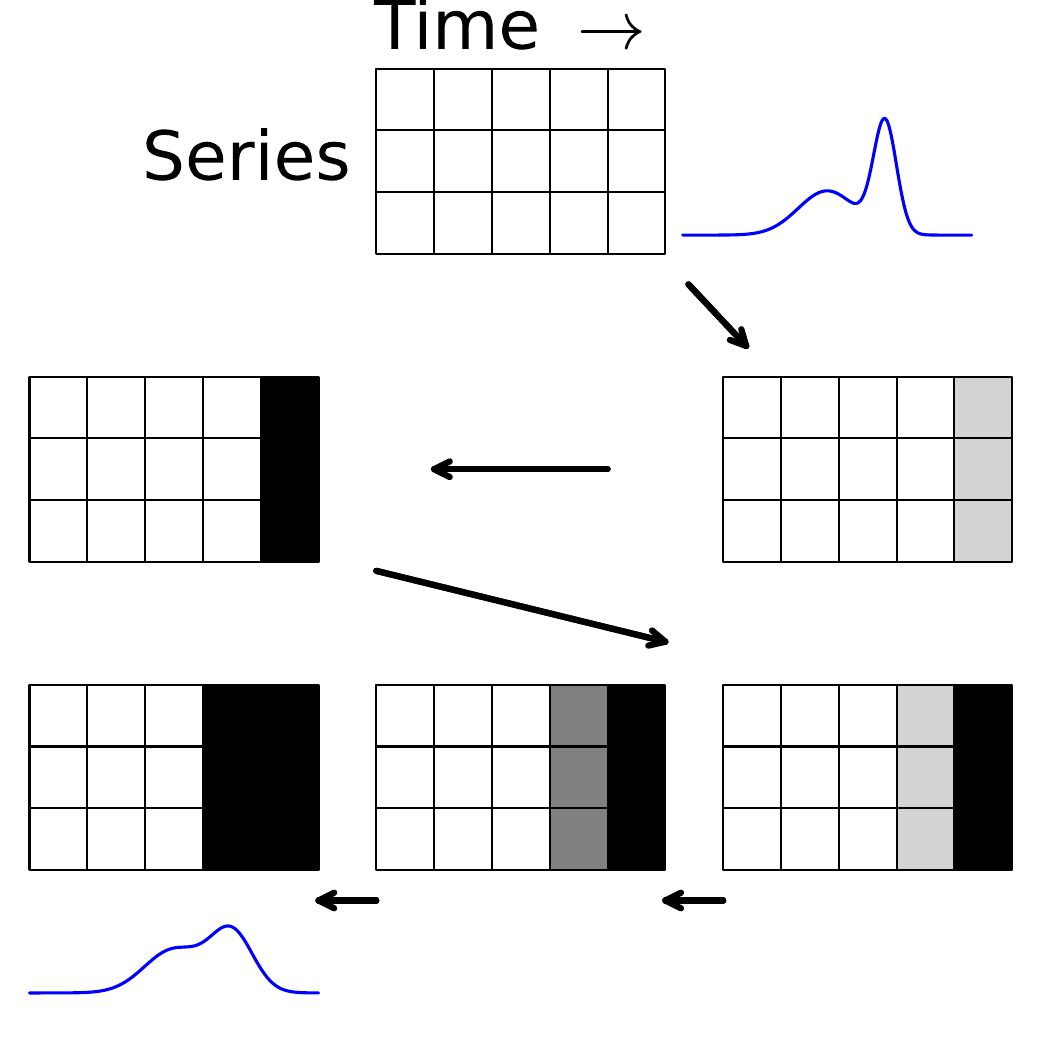} \label{fig:smc-example:leo}}
    }
    \spacingset{1} \caption{
        Different forms of case-deletion for Bayesian hierarchical models and corresponding aSMC-based approximation strategies, based on a single set of baseline (non-case-deleted) samples.
        Each block denotes a conditionally independent unit.
        Black blocks do not contribute to the model likelihood; gray blocks contribute \textit{partially}.
        \textbf{Leave-subset-out} removes a carefully chosen subset of conditionally independent (possibly intra-dependent multivariate) observations. \textbf{Leave-group-out} is a special case.
        \textbf{Leave-end-out} backward-sequentially deletes temporally ordered observations (where each block may be multivariate and non-factorizable).
        \textbf{Across all schemes}, the sampler adaptively constructs auxiliary distributions as necessary to support reliable approximation.
    }
    \label{fig:smc-example}
\end{figure}

Fixing $k$, a design choice lies in the sequence $\gamma_{k,1:L_k} = (\gamma_{k,1}, \ldots, \gamma_{k,L_K})$.
See Figure \ref{fig:smc-example};
with LOO- or LGO-CV (Sections \ref{sec:2:2:1} and \ref{sec:2:2:2}), the goal is to approximate the leave-$g$-out posteriors as the final distribution $p_{g,L_g}$;
with backward-sequential LEO (Section \ref{sec:2:2:3}), the intermediate distributions should contain those induced by leaving the data points backward, and they are also of interest, and not solely the final distribution and its marginal draw.
We briefly explore how to parameterize such structural case-deletions.

\subsubsection{Tempering for joint group deletion} \label{sec:3:2:1}

Tempering is a natural choice to bridge the baseline and target \citep{SwendsenWang1986, MarinariParisi1992, HukushimaNemoto1996, Neal2001}.
Taking the LGO posterior, for instance, we use an augmented likelihood contribution for group $g$ \citep{AgostinelliGreco2013, KallioinenPaanenBurknerVehtari2023} of the form
\begin{equation*}
    \rho_g(n) := p(\bm{y}_g \cond \theta_g, \phi)^{\varphi_g(n)},
\end{equation*}
where $\varphi_g: [0, N_g] \rightarrow [0, 1]$ is a continuous decreasing path such that $\varphi_g(0) = 1$ and $\varphi_g(N_g) = 0$;
we parameterize a \textit{geometric path} of distributions \citep{Neal1993, GelmanMeng1998} from the non-case deleted to leave-$g$-out posterior.
(With LSO, set $N_k=1$.)

A potential drawback is that existing MCMC algorithms, which were effective in targeting the baseline posterior ($n = 0$), may become unsuitable as the invariant kernel within aSMC.
For example, a well-mixing Gibbs sampler that efficiently targets the baseline by exploiting conjugacy might become inapplicable when the power-scaled coefficients in the interior $(0,1)$ induce likelihoods that do not admit conveniently simulatable conjugate priors.
Some exceptions are presented in \cite{KallioinenPaanenBurknerVehtari2023} and in Section \ref{sec:4:yield}.

\subsubsection{Ordered continuous within-group deletions} \label{sec:3:2:2}

A \textit{continuous case deletion} over discrete $i \in \set{1, \ldots, N_g}$ with an augmented likelihood
\begin{equation*}
    \rho_g(n) := \prod_{i=1}^{N_g} p(y_{g,i} \cond \theta_g, \phi)^{\varphi_{g,i}(n)},
\end{equation*}
and the likelihood power-scaling factors $\varphi_{g,i}:[0,N_g] \rightarrow [0,1]$ continuously parameterized by $n$, such as $\varphi_{g,i}(n)
= \min\!\set{
    \max\!\set{
        0,
        i - n
    },
    1
}$,
defines a path such that $\rho_g(0) = p(\bm{y}_g \cond \theta_g, \phi)$ and $\rho_g(N_g) = 1$,
and in particular $\rho_g(n) = \prod_{i=n+1}^{N_g} p(y_{g,i} \cond \theta_g, \phi)$ for integer $n$.

Unlike tempering, this is convenient when an efficient tailored Gibbs sampler exploiting conjugacy is available;
the distribution induced from $n \in \set{1, \ldots, N_g}$ would simply be the posterior with observations $y_{g,i}$ such that $i \leq n$ are left out, which makes it suitable for cases with ordering on the observations (e.g., Figure \ref{fig:smc-example:leo}: by enforcing that such discrete checkpoints are present along the trajectory).
However, the approach is not applicable if conditional independence is violated.

\subsection{Adaptive {mechanisms}} \label{sec:3:3}

Choosing the backward kernels as time reversals in Section \ref{sec:3:1} yields the incremental weight function $w_{k,\ell}(\bm{\Theta}_{\ell-1}) = \gamma_{k,\ell}(\bm{\Theta}_{\ell-1}) / \gamma_{k,\ell-1}(\bm{\Theta}_{\ell-1})$.
Note that this depends only on the previous step $\bm{\Theta}_{\ell-1}$ and not on the $\ell$-th step $\bm{\Theta}_{\ell}$.
This should be leveraged to design user-friendly adaptive {mechanisms} to simplify workflows.

\subsubsection{Automating bridging} \label{sec:3:3:1}

Eliminating the need to explicitly specify the intermediate distributions $(p_{k,1}, \ldots, p_{k,L_k})_{k=1}^{K}$ is advantageous to achieve what is described in Figure \ref{fig:smc-example}, automatically.
\begin{enumerate}[(a)]
    \item In LGO, only marginal draws at the final step from the respective leave-$g$-out posteriors are of interest.
    \item In LEO, draws from the sub-intermediate distributions between the backward-sequentially case deleted distributions are only auxiliary.
\end{enumerate}
The former was suggested in the related work by \cite{BornnDoucetGottaro2010}, though not implemented, and the latter has been touched upon by \cite{BurknerGabryVehtari2020} tangentially.

We implement these as follows.
Given the previous-step particles $\bm{\Theta}_{k,\ell-1}^{(r)} \sim p_{k,\ell-1}(\cdot)$ (now with index $k$) given the case deletion parameter $n_{k,\ell-1}$, we measure (the lack of) weight diversity by the effective sample size \citep[ESS:][]{KongLiuWong1994},
\begin{equation*}
    \operatorname{ESS}_{k,\ell} := {1 \over \sum_{r=1}^R (W_{k,\ell}^{(r)})^2},
    \qquad
    W_{k,\ell}^{(r)} = {w_{k,\ell}^{(r)} \over \sum_{r=1}^R w_{k,\ell}^{(r)}},
    \qquad
    w_{k,\ell}^{(r)}
    = {\gamma_{k,\ell}(\bm{\Theta}_{k,\ell-1}^{(r)}) \over \gamma_{k,\ell-1}(\bm{\Theta}_{k,\ell-1}^{(r)})}.
\end{equation*}
Asymptotic connections between ESS and $\chi^2$-divergence between the target and proposal distributions have been discussed in \cite{AgapiouPapaspiliopoulosSanz-AlonsoStuart2017}.
Under the case deletion parameterizations discussed in Section \ref{sec:3:2}, the unnormalized weights can then be explicitly expressed as a function of the power coefficient in $[0,1]$.
We can solve for $n_{k,\ell} \in (0, N_k]$ (or up to the pre-determined sub-intermediate \textit{checkpoint} $n_{k,\ell+1} \leq N_k$ for LEO) to determine the next target distribution such that ESS meets a specified threshold, as ESS decreases in $n_{k,\ell}$ between the sub-intermediates;
safe root-finding algorithms such as the bisection or Brent's method can be used \citep{BeskosJasraKantasThiery2016}. 
See also \cite{CornebiseMoulinesOlsson2008}, \cite{JasraStephensDoucetTsagaris2011}, and \cite{DelMoralDoucetJasra2012}.

\subsubsection{Diagnostics} \label{sec:3:3:2}

It is unclear \textit{a priori} whether subset deletion necessitates particle rejuvenation via an invariant kernel.
For instance, when $N_g=1$, the excluded observation may induce minimal posterior shift, and the sampler may proceed without intermediate steps (i.e., $L_k = 1$).
In such cases, given that the invariant kernel constitutes the most computationally intensive component, IS may suffice;
an option is to invoke the invariant kernel selectively contingent on diagnostics.
Since the baseline draws from step $L_k-1 = 0$ can be used to compute the next step ($L_k=1$) importance weights, the reliability of the IS estimate can immediately be assessed prior to applying the invariant kernel, using any suitable metric.
We use the generalized Pareto $\hat{k}$ diagnostic (e.g., $\hat{k} > 0.7$ as recommended by \citealp{VehtariSimpsonGelmanYaoGabry2024} and \citealp{Millar2018}), along with the ESS criterion.
If both indicate stability, we proceed with the fast(er) {Pareto-smoothed IS (PSIS)}.
Otherwise, the invariant kernel is triggered to rejuvenate the particles;
the approach embeds the conventional heuristic of \textit{re-running MCMC} only when necessary \citep[e.g.,][]{BurknerGabryVehtari2020} into the aSMC sampler.

\subsection{Choice of estimands} \label{sec:3:4}

The idea behind the sequential approximation is motivated by the identity in \eqref{eq:smc:radon-nikodym}.
Taking \eqref{eq:lgo-lpd-est:is} for instance, estimating $\ell^{(\text{LGO})}$ could be considered as a special case in which the target functions are defined as
\begin{equation*}
    f_g(\bm{\Theta})
    := p(\bm{y}_g^* = \bm{y}_g \cond \theta_g, \phi),
\end{equation*}
and the final approximating quantity $\hat{\ell}^{(\text{LGO})}$ is obtained as a sum of the approximate logarithmic scores $\hat{\ell}_g^{(\text{LGO})}$.
We emphasize, from an algorithmic point of view (the focus of this paper) that the case-deletion scheme and the selection of an estimand can be treated as distinct and independent operations under the user's control.
In light of this, we advocate choosing the estimand to reflect hypothetical data replications and the specific aspects of the out-of-sample generalization that are most relevant \citep{GelmanHwangVehtari2014}.

It is also often preferable to select scoring rules familiar to subject-matter experts, over relying solely on the predictive densities as default measures whose differences or variabilities may be difficult to interpret.
For example, in Bayesian econometric time-series applications, brute-force LEO is often used to evaluate Bayesian point forecasts relative to frequentist counterparts using standard error metrics \citep[e.g.,][]{FaustWright2013}, or density forecasts via log predictive likelihood \citep[e.g.,][]{KoopKorobilisPettenuzzo2019}.

That predictive density is not necessarily the default highlights that out-of-sample model checking need not center on predicting new observations.
Beyond predictive densities, other targets of interest may be the shared parameter $f_g(\bm{\Theta}) := \phi$ and the group-specific parameter $f_g(\bm{\Theta}) := \theta_g$.
Extrapolation to a new group $G+1$ may be considered via leave-$g$-out-integrable $f_g(\bm{\Theta}, \bm{y}_{G+1}^*, \theta_{G+1})$ by operating on the augmented posterior with a new group $G+1$ which admits the original leave-$g$-out posterior as its marginal.

Finally, we note the asymptotic equivalence of Bayesian CV and the widely available information criterion \cite[WAIC:][]{Watanabe2010}, as well as the correspondence of different forms of WAIC to different forms of LOO-CV and LGO-CV estimands \citep{GelmanHwangVehtari2014, MerkleFurrRabe-Hesketh2019}.
With regard to the choice of target functions, we further acknowledge discussions that are generally in favor of the use of marginal likelihoods over conditional likelihoods, due to improved numerical stability and accurate approximation \citep[of WAIC:][]{LiQiuZhangFeng2015},
and the fact that marginal measures align better with the regularity condition for the asymptotic equivalence to hold \citep{Millar2018}.
The target function may then be appropriately selected to target these marginal estimands, provided that marginal likelihoods can be evaluated; a further step may need to be implemented to approximate the integral, such as by quadrature \citep[Appendix C]{MerkleFurrRabe-Hesketh2019}.

\subsection{Summary and relation to existing approach}

Algorithm \ref{alg:2} details the adaptive approach.
For clarity, a diagrammatic illustration of the sampler with three CV schemes is {provided in Figure \ref{fig:smc-example}}.

\begin{center}
\begin{minipage}{\linewidth}
\spacingset{1.1}
\begin{algorithm}[H]
\SetAlgoLined
    \KwIn{MCMC draws $\bm{\Theta}^{(1)}, \ldots, \bm{\Theta}^{(R)} \sim p(\bm{\Theta} \cond \bm{y}_{1:G})$
    }
    \KwResult{$\hat{\ell}_{1:K}, \hat{\ell}$}
        \For{$k = 1, \ldots, K$ in parallel}{
            Initialize index $\ell \gets 0$ \;
            Initialize case deletion parameter $n_{\ell=0} \gets 0$ \;
            Initialize particles $(\bm{\Theta}_{0}^{(r)}, W_{0}^{(r)}) \gets (\bm{\Theta}^{(r)}, 1/R)$  \tcp*{Index $k$ omitted}
            \While{$n_{\ell-1} < N_k$}{
                $\ell \gets \ell + 1$ \;
                Solve $n_{\ell} \in (n_{\ell - 1}, N_k]$ \tcp*{Section \ref{sec:3:3:1}}
                Compute $(W_{\ell}^{(1)}, \ldots, W_{\ell}^{(R)})$ from $n_{\ell}$ \tcp*{Equation \ref{eq:smc:weight}}
                \uIf{$n_{\ell} < N_k$}{
                    Deduce $\gamma_{k,\ell}$ from $n_{\ell}$ \tcp*{Section \ref{sec:3:2}}
                    \nl Compute $W_{\ell}^{(r)} \propto w_{\ell}^{(r)} = (\gamma_{k,\ell} / \gamma_{k,\ell-1})(\bm{\Theta}_{\ell-1}^{(r)})$ \label{alg:mcmc:begin} \;
                    \nl $A_{\ell}^{(r)} \sim \texttt{Resample}(W_{\ell}^{(1)}, \ldots, W_{\ell}^{(R)})$ \;
                    \nl $\bm{\Theta}_{\ell}^{(r)} \sim \mathcal{K}_{k,\ell}(~\cdot \leftarrow \bm{\Theta}_{\ell-1}^{(A_{\ell}^{(r)})})$ in parallel \tcp*{Invariant kernel \eqref{sec:3:1}}
                    \nl $(\bm{\Theta}_{\ell}^{(r)}, W_{\ell}^{(r)}) \gets (\bm{\Theta}_{\ell}^{(r)}, 1/R)$ \label{alg:mcmc:end} \;
                }
                \Else{
                    
                    $\hat{k}, (\widehat{W}_{\ell}^{(1)}, \ldots, \widehat{W}_{\ell}^{(R)}) \gets \texttt{ParetoSmooth}(W_{\ell}^{(1)}, \ldots, W_{\ell}^{(R)})$ \;
                    \uIf{$\hat{k} < 0.7$}{
                        $(\bm{\Theta}_{\ell}^{(r)}, W_{\ell}^{(r)}) \gets (\bm{\Theta}_{\ell-1}^{(r)}, \widehat{W}_{\ell}^{(r)})$ \tcp*{Optional (\ref{sec:3:3:2})}
                    }
                    \Else{
                        Rejuvenate $\bm{\Theta}_\ell^{(r)}$ in parallel \tcp*{as in lines \ref{alg:mcmc:begin} to  \ref{alg:mcmc:end}}
                    }
                    $L_k \gets \ell$ \tcp*{$n_{\ell} = N_k$; end loop}
                }
                \medskip
                $\hat{\ell}_{k,\ell} \gets \log \sum_{r=1}^R W_{\ell}^{(r)} f_k(\bm{y}_{\mathcal{I}_k}, \bm{\Theta}_{\ell}^{(r)})$  \tcp*{Sections (\ref{sec:2:2}; \ref{sec:3:4})}
            }
            \medskip
            $\hat{\ell}_{k} \gets \hat{\ell}_{k,L_k}$ \;
            \tcp{Further continue for LEO}
        }
        $\hat{\ell} \gets \texttt{Aggregate}(\hat{\ell}_{1:K}) \overset{\text{e.g.}}{=}\sum_{k=1}^K \hat{\ell}_{k}$ \;
    {\bf return} $\hat{\ell}_{1:K}, \hat{\ell}$
    \spacingset{1} \caption{\texttt{Adaptive SMC for structural Bayesian cross validation}}
    \label{alg:2}
\end{algorithm}
\end{minipage}
\end{center}

The proposed approach can be viewed as a direct extension of previous works \citep{GelfandDey1994, Peruggia1997, EpifaniMacEachernPeruggia2008, BornnDoucetGottaro2010, VehtariGelmanGabry2017, BurknerGabryVehtari2020} using (PS)IS for approximate LOO-CV.
The algorithm {complements these works in that} we operate on a continuum of distributions which are easier to approximate.
The auxiliary intermediate distributions are determined fully automatically to streamline the workflow.
The selection of PSIS and MCMC re-runs (in the sense that the MCMC kernel is applied) is guided via the ESS criterion \citep{KongLiuWong1994, AgapiouPapaspiliopoulosSanz-AlonsoStuart2017} and partly the generalized Pareto shape diagnostic \citep{VehtariSimpsonGelmanYaoGabry2024} to minimize unnecessary MCMC re-runs where appropriate.
When the MCMC kernel is invoked, it serves as a design-efficient alternative to fully re-running MCMC, under reasonable parallel resources (as we demonstrate in Section \ref{sec:4}), since it enables particle-wise parallelization without extensive burn-in or full-chain regeneration.
We thereby extend the non-adaptive SMC approach {of \cite{BornnDoucetGottaro2010} for LOO-CV} specifically on Bayesian LASSO \citep{ParkCasella2008},
and then \cite{VehtariGelmanGabry2017} and \cite{BurknerGabryVehtari2020} to subsume the workflow of MCMC re-runs as an efficient systematic component of the sampler.
The sampler also supports various CV designs, including
LGO \citep{MerkleFurrRabe-Hesketh2019, LiuRue2023},
LEO \citep{BurknerGabryVehtari2020}, and
LSO (e.g., $K$-fold).


\section{Applications} \label{sec:4}

We illustrate the proposed approach using real data examples.
Throughout the examples, unless otherwise noted,
we obtain 1000 samples from the non-case-deleted posterior via 4000 iterations of the dynamic \citep{HoffmanGelman2014} Hamiltonian Monte Carlo (HMC), discarding the initial 1000 and applying a thinning factor of 3 upon inspecting empirical autocorrelation.
{We then use HMC as the aSMC invariant kernel with 1--3 iterations (5 for Gibbs) chosen in consideration of differences in autocorrelation, which is available from the baseline draw.}
The aSMC sampler uses the 1000 resulting samples as its initial high-quality marginal draw and operates with an ESS ratio threshold of 0.5 (which is typical).
{We note that, since the MCMC kernel is applied per particle, the run time increases proportionally with the number of kernel applications;
it is therefore prudent to set the number of iterations at a minimal sufficient level for effective low-autocorrelation particle rejuvenation.
That said, because rejuvenations are independent particle-wise, distributed computing can substantially reduce run time, and in principle down to that required for a single-particle rejuvenation provided the parallel resources.
We employ 8-thread multi-threading unless otherwise noted.
Moreover, to further promote efficiency, we advocate repurposing information already available from baseline dynamic HMC posterior draws at no additional cost
(e.g., autocorrelations, as above, and covariances for the HMC mass matrix).}
\texttt{Julia} code is available at \href{https://github.com/geonhee619/aSMC-BayesCV}{https://github.com/geonhee619/aSMC-BayesCV}.

\subsection{Hierarchical example} \label{sec:4:radon}

\subsubsection{Radon exposure multilevel regression}

This section considers a hierarchical example in which group sizes vary substantially with within-group observation counts $N_g$ ranging from 1 to 116.
Following \cite{VehtariGelmanGabry2017}, consider the Bayesian multilevel {model that describes the measurement} of radon in households in Minnesota,
\begin{align*}
    y_i
        \ind
        \mbox{normal}(\bm{x}_i^\tr \bm{\beta}_{g[i]}, \sigma), \qquad
    \bm{\beta}_g
        \ind \mbox{MVN}\!\bracket{
            \bm{\Gamma} \bm{u}_g,
            \bm{\Sigma}
        },
\end{align*}
where
$y_i$ represents the measurements of the radon concentration on a logarithmic scale.
$\mbox{normal}(\mu, \sigma)$ is the univariate normal distribution with location and scale $(\mu, \sigma)$.
The measurement-level predictor $\bm{x}_i = (1, x_i)^\tr$ includes an intercept and a binary indicator $x_i$ set to one if the measurement was taken on the first floor and zero if taken in the basement.
The group- or county-level predictor $\bm{u}_g = (1, u_g)^\tr$ is observed, where $u_g$ is the soil uranium level in county $g$ also on a logarithmic scale.
The indices run over $i \in \set{1, \ldots, N}$ and $g \in \set{1, \ldots, G}$, where $N = 919$ denotes the number of observations and $G = 85$ represents the number of counties.
For a more complete description of the data and setup, we refer to \cite{PriceNeroGelman1996} and \cite{GelmanHill2006}.

\subsubsection{Leave-group-out cross validation}

\begin{figure}[h]
     \centering\spacingset{1}
     \subfloat[MCMC]{
        \tikz{
             \node[obs] (Theta_1) {$\bm{\Theta}^{(1)}$};
             \node[obs, below=of Theta_1] (Theta_2) {$\bm{\Theta}^{(2)}$};
             \node[obs, below=of Theta_2] (Theta_R) {$\bm{\Theta}^{(R)}$};
             \edge[]{Theta_1}{Theta_2};
             \edge[]{Theta_2}{Theta_R};
    
             \node[latent, right=of Theta_1, xshift=-0.5cm] (Theta_1_g) {$\bm{\Theta}_g^{(1)}$};
             \node[latent, below=of Theta_1_g] (Theta_2_g) {$\bm{\Theta}_g^{(2)}$};
             \node[latent, below=of Theta_2_g] (Theta_R_g) {$\bm{\Theta}_g^{(R)}$};
             \edge[]{Theta_1_g}{Theta_2_g};
             \edge[]{Theta_2_g}{Theta_R_g};
            
            \plate [xshift=0cm] {plate} {(Theta_1_g) (Theta_R_g)} {$g \in [G]$};
         }
     } \quad\quad
     \subfloat[IS]{
        \tikz{
             \node[obs] (Theta_1) {$\bm{\Theta}^{(1)}$};
             \node[obs, below=of Theta_1] (Theta_2) {$\bm{\Theta}^{(2)}$};
             \node[obs, below=of Theta_2] (Theta_R) {$\bm{\Theta}^{(R)}$};
             \edge[]{Theta_1}{Theta_2};
             \edge[]{Theta_2}{Theta_R};
    
             \node[latent, right=of Theta_1, xshift=-0.5cm] (Theta_1_1) {$\bm{\Theta}^{(1)}_{g}$};
                \edge{Theta_1}{Theta_1_1};
    
            \node[latent, right=of Theta_2, xshift=-0.5cm] (Theta_2_1) {$\bm{\Theta}^{(2)}_{g}$};
                \edge{Theta_2}{Theta_2_1};
    
            \node[latent, right=of Theta_R, xshift=-0.5cm] (Theta_R_1) {$\bm{\Theta}^{(R)}_{g}$};
                \edge{Theta_R}{Theta_R_1};
    
            \edge{Theta_1}{Theta_1_1, Theta_2_1, Theta_R_1};
            \edge{Theta_2}{Theta_1_1, Theta_2_1, Theta_R_1};
            \edge{Theta_R}{Theta_1_1, Theta_2_1, Theta_R_1};
            
            \plate [xshift=-0.1cm] {plate} {(Theta_1_1) (Theta_R_1)} {$g \in [G]$};
         }
     } \quad\quad
     \subfloat[aSMC]{
         \tikz{
             \node[obs] (Theta_1) {$\bm{\Theta}^{(1)}$};
             \node[obs, below=of Theta_1] (Theta_2) {$\bm{\Theta}^{(2)}$};
             \node[obs, below=of Theta_2] (Theta_R) {$\bm{\Theta}^{(R)}$};
             \edge[]{Theta_1}{Theta_2};
             \edge[]{Theta_2}{Theta_R};
    
             \node[obs, right=of Theta_1, xshift=-0.5cm] (Theta_1_1) {$\bm{\Theta}^{(1)}_{g,0}$};
                \edge{Theta_1}{Theta_1_1};
             \node[latent, right=of Theta_1_1] (Theta_1_2) {$\bm{\Theta}^{(1)}_{g,1}$};
                \edge{Theta_1_1}{Theta_1_2};
             \node[latent, right=of Theta_1_2] (Theta_1_L) {$\bm{\Theta}^{(1)}_{g,L_g}$};
                \edge[dashed]{Theta_1_2}{Theta_1_L};
    
            \node[obs, right=of Theta_2, xshift=-0.5cm] (Theta_2_1) {$\bm{\Theta}^{(2)}_{g,0}$};
                \edge{Theta_2}{Theta_2_1};
             \node[latent, right=of Theta_2_1] (Theta_2_2) {$\bm{\Theta}^{(2)}_{g,1}$};
                \edge{Theta_2_1}{Theta_2_2};
             \node[latent, right=of Theta_2_2] (Theta_2_L) {$\bm{\Theta}^{(2)}_{g,L_g}$};
                \edge[dashed]{Theta_2_2}{Theta_2_L};
    
            \node[obs, right=of Theta_R, xshift=-0.5cm] (Theta_R_1) {$\bm{\Theta}^{(R)}_{g,0}$};
                \edge{Theta_R}{Theta_R_1};
             \node[latent, right=of Theta_R_1] (Theta_R_2) {$\bm{\Theta}^{(R)}_{g,1}$};
                \edge{Theta_R_1}{Theta_R_2};
             \node[latent, right=of Theta_R_2] (Theta_R_L) {$\bm{\Theta}^{(R)}_{g,L_g}$};
                \edge[dashed]{Theta_R_2}{Theta_R_L};
    
            \edge{Theta_1_1}{Theta_1_2, Theta_2_2, Theta_R_2};
            \edge{Theta_2_1}{Theta_1_2, Theta_2_2, Theta_R_2};
            \edge{Theta_R_1}{Theta_1_2, Theta_2_2, Theta_R_2};
    
            \edge[dashed]{Theta_1_2}{Theta_1_L, Theta_2_L, Theta_R_L};
            \edge[dashed]{Theta_2_2}{Theta_1_L, Theta_2_L, Theta_R_L};
            \edge[dashed]{Theta_R_2}{Theta_1_L, Theta_2_L, Theta_R_L};
            
            \plate [xshift=-0.1cm] {plate} {(Theta_1_1) (Theta_R_L)} {$g \in [G]$ groups};
         }
    } \spacingset{1} \caption{
        Diagrammatic comparison of three strategies for approximate \textit{leave-group-out cross-validation}.
        {The solid arrow ($\rightarrow$) represents the algorithmic input-output relationship.}
        The draws $\bm{\Theta}^{(1)}, \ldots, \bm{\Theta}^{(R)} \sim p(\cdot \cond \bm{y}_{1:G})$ are from the baseline non-case-deleted posterior distribution, MCMC, say.
        {At the post-MCMC stage, we seek draws from the $\bm{y}_g$-deleted posterior using what is available (gray), to approximate latent quantities not directly accessible (white)}, for $g \in [G] := \set{1, \ldots, G}$.
        (a) MCMC constructs Markov chains that leave the $G$ leave-$g$-out posteriors invariant (with appropriate burn-in and thinning).
        (b) IS exploits baseline samples as input to produce one-step weighted approximations $p(\bm{\Theta}_g \cond \bm{y}_{-g}) \approx \sum_{r=1}^R W^{(r)} \delta_{\bm{\Theta}^{(r)}}(\bm{\Theta}_g)$.
        (c) aSMC yields multi-step weighted approximations $\sum_{r=1}^R W_{L_g}^{(r)} \delta_{\bm{\Theta}_{L_g-1}^{(r)}}(\cdot)$.
        {The dashed arrow ($\dashrightarrow$) indicates that $L_g \geq 1$ is determined adaptively and is thus unknown \textit{a priori}.}
    } \label{fig:smc}
\end{figure}

We consider LGO-CV; see Figures \ref{fig:smc-example:lgo} and \ref{fig:smc}.
The parameters in the non-case deleted posterior would be $\bm{\Theta} = (\bm{\beta}_{1:G}, \sigma, \bm{\Gamma}, \bm{\Sigma})$ partitioned with $\phi = (\bm{\Gamma}, \bm{\Sigma}, \sigma)$ and $\theta_g = \bm{\beta}_g$ according to the notation in Section \ref{sec:2}.
The unnormalized importance weights are
\begin{equation*}
    w_h
    = \prod_{i : g[i] = h}
    \mbox{normal}(y_i \cond \bm{x}_i^\tr \bm{\beta}_{h}, \sigma)^{-1}.
\end{equation*}
Although the dataset, with 919 observations and 85 counties, is not excessively large, a single run of MCMC to obtain draws from the non-case-deleted posterior takes approximately 15 minutes.
Na\"ively extending this to compute the LGO estimands could result in a total run time of up to 21 hours.
Reducing inefficiencies lets applied modelers focus their workflow on more meaningful diagnostic checks and model expansion.

\begin{figure}
    \centering\spacingset{1}
    \includegraphics[width=\textwidth]{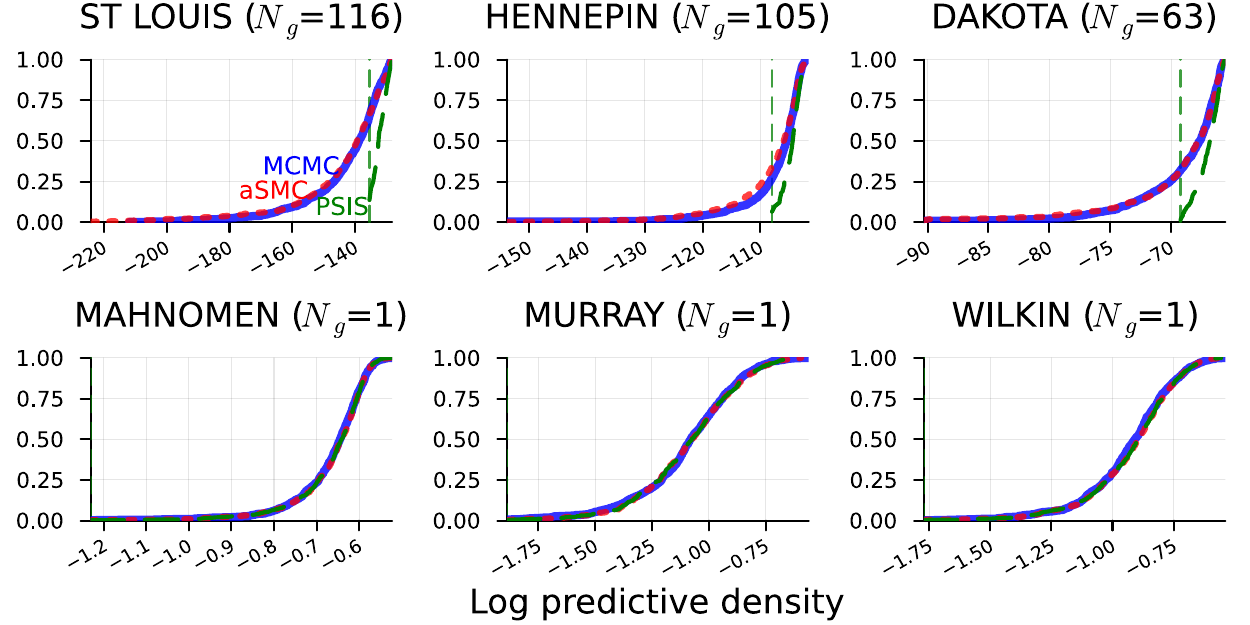}
    \spacingset{1} \caption{
        \textit{Hierarchical example}.
        Comparison of posterior distribution function approximations of LGO log predictive likelihood.
        \textbf{Note}:
        Displayed in the top row are the three most populous groups in the data; the bottom row shows the three least.
        The solid line corresponds to the MCMC approximation.
        The dotted vertical line indicates the point at which the PSIS approximation ends.
    }
    \label{fig:radon:compare}
\end{figure}

Figure \ref{fig:radon:compare} first shows the empirical distribution of log predictive likelihood from LGO posteriors, comparing the top and bottom three groups ranked by within-group sample size.
Note that we are taking the logarithm; this is to facilitate visible comparison.
Treating MCMC as reference, aSMC produces approximations highly close to those of MCMC, especially when the within-group observations $N_g$ is high, where IS {estimators fail to} approximate the tails of the group-deleted posteriors.
Results are nearly identical for groups with a single observation, which is expected.

\begin{figure}
    \centering\spacingset{1}
    \includegraphics[width=\textwidth]{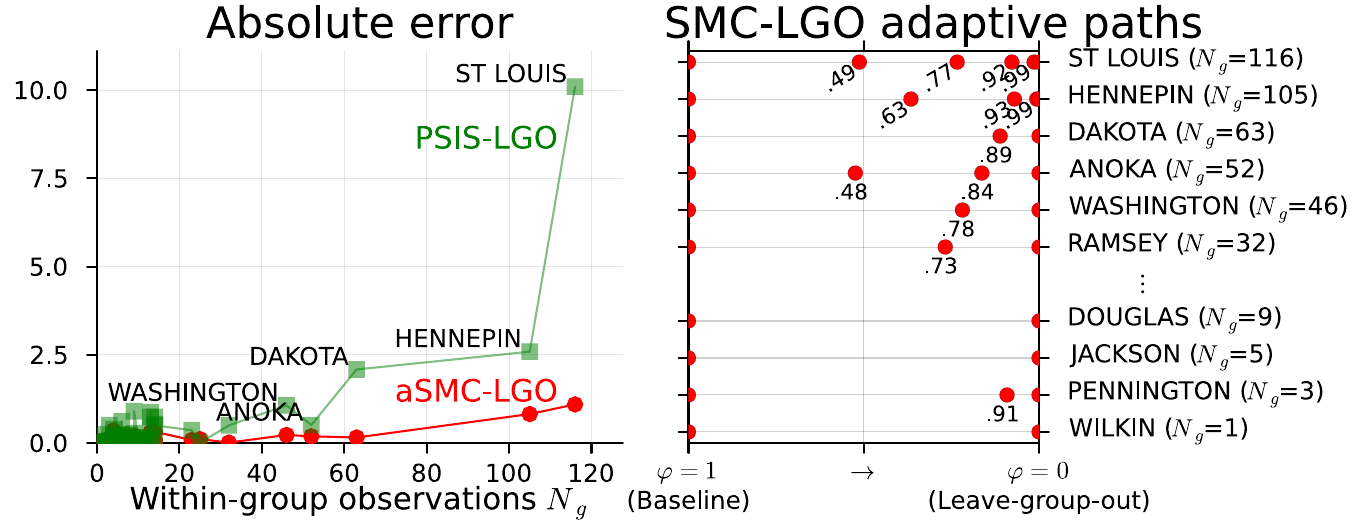}
    \spacingset{1} \caption{
        \textbf{Left}: Error comparison for LGO log predictive likelihood estimates, benchmarked against the brute-force strategy of re-running MCMC for each group (county).
        \textbf{Right}: Realized trajectories of distributions automatically determined by aSMC.
        \textbf{Note}: Annotated numbers denote the determined likelihood-contribution power coefficients.
        The bisection method is used to perform the root-finding step (see Section \ref{sec:3:3:1}).
        Counties are ordered by the number of within-group observations $N_g$ {(in parentheses)}.
    }
    \label{fig:radon:compare-error}
\end{figure}

Figure \ref{fig:radon:compare-error} plots the absolute error against within-group observations $N_g$, using the MCMC-based estimates as reference.
On the low end, the two estimators produce practically identical results.
As $N_g$ increases, the quality of PSIS estimates degrades, while aSMC maintains a more reliable performance.

Figure \ref{fig:radon:compare-error} also visualizes the paths.
More intermediate distributions are typically configured for groups with large within-group sample sizes, and often none for smaller ones.
Adaptive bridging therefore streamlines the workflow by automating both the design of the sequential path and the decision of when to invoke the MCMC kernel.

\begin{figure}
    \centering\spacingset{1}
    \includegraphics[width=0.9\textwidth]{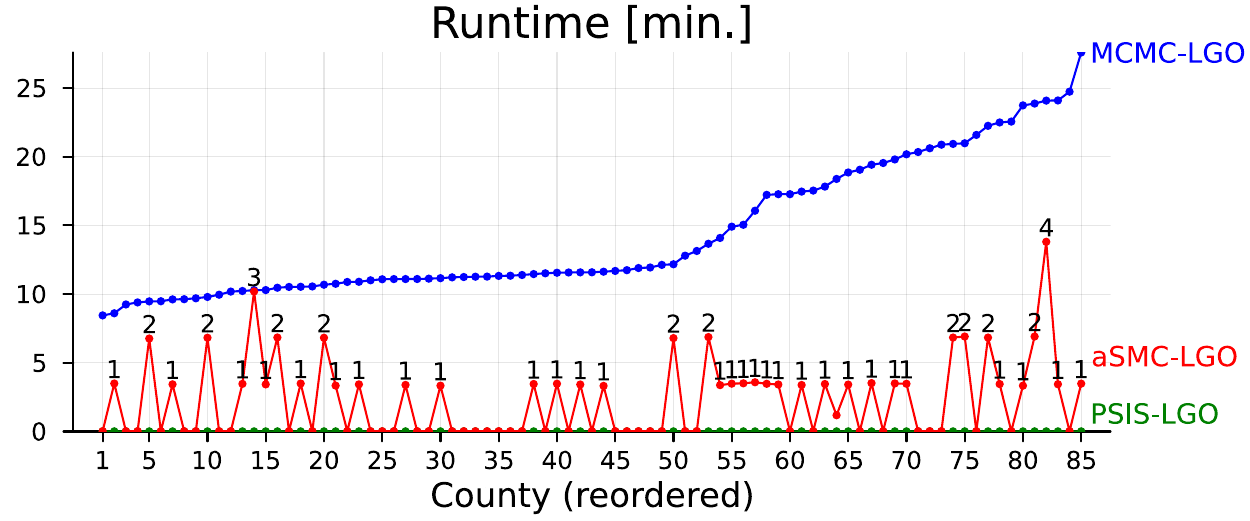}
    \spacingset{1} \caption{
        Comparison of run time.
        Counties are ordered by MCMC run time.
        The numbers annotated on the aSMC run times are the total number of intermediate distributions excluding the baseline and final LGO distribution.
        MCMC run time variation arises from HMC adaptation tailored to each LGO posterior geometry, performed during the burn-in phase.
    }
    \label{fig:radon:runtime}
\end{figure}

Figure \ref{fig:radon:runtime} then compares the run time of aSMC versus re-running MCMC.
As operations can be parallelized across counties {for both approaches}, we compare the run time per county.
aSMC is faster than MCMC for all counties.
The run time is negligible for counties without intermediate distributions, in which case the procedure reduces to PSIS, which is fast.
In cases with at least one intermediate distribution, run time increases with the number of distributions but remains faster than MCMC re-runs, yielding substantially faster total run time with comparable estimates.

\subsection{Time-series example} \label{sec:4:yield}

This section illustrates model validation of a Bayesian state-space model using the backward-sequential LEO scheme (Section \ref{sec:2:2:3}).

\subsubsection{Yield curve forecasting}

Forecasting the term structure of interest rates plays a central role in macroeconomic analysis, as the yield spread has consistently demonstrated predictive ability for future macroeconomic conditions.
\cite{EstrellaHardouvelis1991} documented the yield spread, the difference between the rates of the 10-year Treasury bond and the three-month Treasury bill, as an effective predictor of future macroeconomic variables.
\cite{HamiltonKim2002} highlighted the predictive capacity of the yield spread for real GDP growth.

\cite{DieboldLi2006} introduced a time-varying factor representation of the term structure of interest rates as the dynamic Nelson--Siegel (DNS) model.
The DNS models the yield for a specific maturity $\tau$ as
\begin{equation*}
    \mu_t(\tau)
        = \beta^{(l)}_t
        + \beta^{(s)}_t \bracket{1 - \exp(-\lambda_t \tau) \over \lambda_t \tau}
        + \beta^{(c)}_t \bracket{{1 - \exp(-\lambda_t \tau) \over \lambda_t \tau} - \exp(-\lambda_t \tau)},
\end{equation*}
where factors $\bm{\beta}_t = (\beta_t^{(l)}, \beta_t^{(s)}, \beta_t^{(c)})^\tr$ represent the time-varying level (long-term yields), slope (short- and long-term spread), and curvature (midterm hump).
$\lambda_t$ controls the exponential decay rate of the curve.
The factors evolve smoothly as
\begin{equation*}
    \bm{\beta}_t = \bm{\beta}_{t-1} + \bm{\varepsilon}_t^{(\beta)}, \qquad 
    \bm{\varepsilon}_t^{(\beta)} \iid \mbox{MVN}(\bm{0}, \bm{\Sigma}^{(\beta)}).
\end{equation*}

Bayesian extensions to the DNS model have since been proposed \citep[e.g.,][]{LauriniHotta2010, Abanto-ValleLachosGhosh2012}.
We focus on assessing the Bayesian rendition given data of the monthly yield from Japanese government bonds.
The dataset spans from September 1999 to January 2024 and includes maturities $\tau \in \mathcal{T} = \set{2, 5, 10, 20, 30}$, which was the longest available time frame with complete data for these maturities at the time of analysis.

We complete the Bayesian model specification first by the measurement equation,
\begin{equation*}
    \bm{y}_t
        = \begin{bmatrix}
            \mu_t(\tau_1) \\
            \vdots \\
            \mu_t(\tau_K)
        \end{bmatrix} + \bm{\varepsilon}_t^{(y)}, \qquad
     \bm{\varepsilon}_t^{(y)}
        \iid \mbox{MVN}(\bm{0}, \bm{\Sigma}^{(y)}),
\end{equation*}
where $\bm{y}_t$ denotes the yields observed across maturities $\mathcal{T}$, with $K$ indicating the number of maturities.
$\bm{\varepsilon}_t^{(y)}$ captures measurement noise.
The priors we impose are:
the initial state $\bm{\beta}_0 = (\beta_0^{(l)}, \beta_0^{(s)}, \beta_0^{(c)})^\tr \sim \mbox{MVN}(\bm{m} = \bm{0}, \bm{P}^{-1} = 10\bm{I}_3)$,
noise covariances $\bm{\Sigma}^{(y)} \sim \textsc{IW}(\nu_0^{(y)} = 2K, \bm{S}_0^{(y)} = \bm{I}_K)$, $\bm{\Sigma}^{(\beta)} \sim \textsc{IW}(\nu_0^{(\beta)} = 2(3), \bm{S}_0^{(\beta)} = \bm{I}_3)$
(where $\textsc{IW}(\nu_0, \bm{S}_0)$ is the {inverse Wishart distribution} with degrees of freedom $\nu_0$ and scale matrix $\bm{S}_0$).
{We set $\lambda_t = 0.0609$ \citep{DieboldLi2006} to simplify estimation, as the main focus of this section is model validation.}
Under the notation in Section \ref{sec:2}, we have
$\phi = (\bm{\Sigma}^{(\beta)}, \bm{\beta}_{0:T}, \bm{\Sigma}^{(y)})$,
with $G = 1$ (e.g., single country) and index $i = t$; each conditionally independent $y_{(g=1),t} := \bm{y}_t$ is treated as $K$-variate.

\subsubsection{Backward-sequential LEO}

The LEO scheme sequentially deletes the final dependent observation, as illustrated in Figure \ref{fig:smc-example:leo}.
The corresponding case-deleted posteriors are deterministically injected as intermediate distributions, and between these, sub-intermediate distributions are further introduced adaptively by the sampler.
Continuous case deletions are applied (see Section \ref{sec:3:2:2}), as the sub-intermediate models admit fast full conditional Gibbs sampling, executed in parallel across particles via 12-thread multi-threading;
see Appendix for further details.
To obtain 1000 baseline particles, we ran 12000 iterations of the Gibbs sampler with 2000 burn-in samples and a thinning factor of 10 (in consideration of autocorrelation).

\begin{figure}
    \centering
    \includegraphics[width=0.8\linewidth]{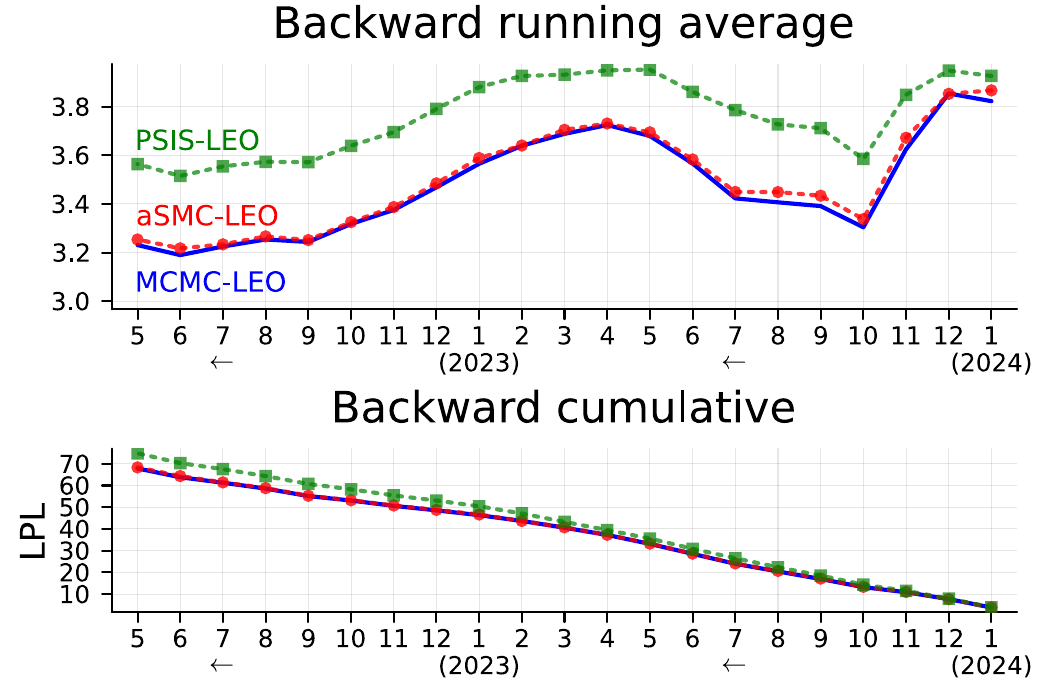}
    \spacingset{1} \caption{
        \textit{Time-series example}.
        Comparison of log predictive likelihood (LPL) approximation.
        Both are computed \textit{backwards} as observations are sequentially deleted backwards.
        {The leftmost value corresponds to the estimate (a) using the least amount of in-sample data points from the model's perspective (as subsequent data points are deleted) and (b) the most amount of data points from perspective of one-step-ahead log predictive likelihood estimation (as the deleted cases are treated as out-of-sample data)}.
    }
    \label{fig:yield-lpd}
\end{figure}

Figure \ref{fig:yield-lpd} illustrates the cumulative and running-average log predictive likelihoods for one-step-ahead forecasts.
The target function is the log predictive likelihood, to facilitate visual comparison.
The cumulative likelihoods are computed in reverse, consistent with the backward-sequential LEO scheme;
the running average yields a backward estimate of the one-step-ahead log predictive likelihood.

Focusing on the running average, the aSMC sampler closely approximates the estimates obtained via the brute-force MCMC approach.
IS estimates degrade with longer horizons and therefore more intensive deletions;
the diagnostic measure $\hat{k}$ is almost always above 0.7.
aSMC sampler avoids this degradation by rejuvenating the particles where appropriate.

\begin{figure}
    \centering
    \includegraphics[width=\linewidth]{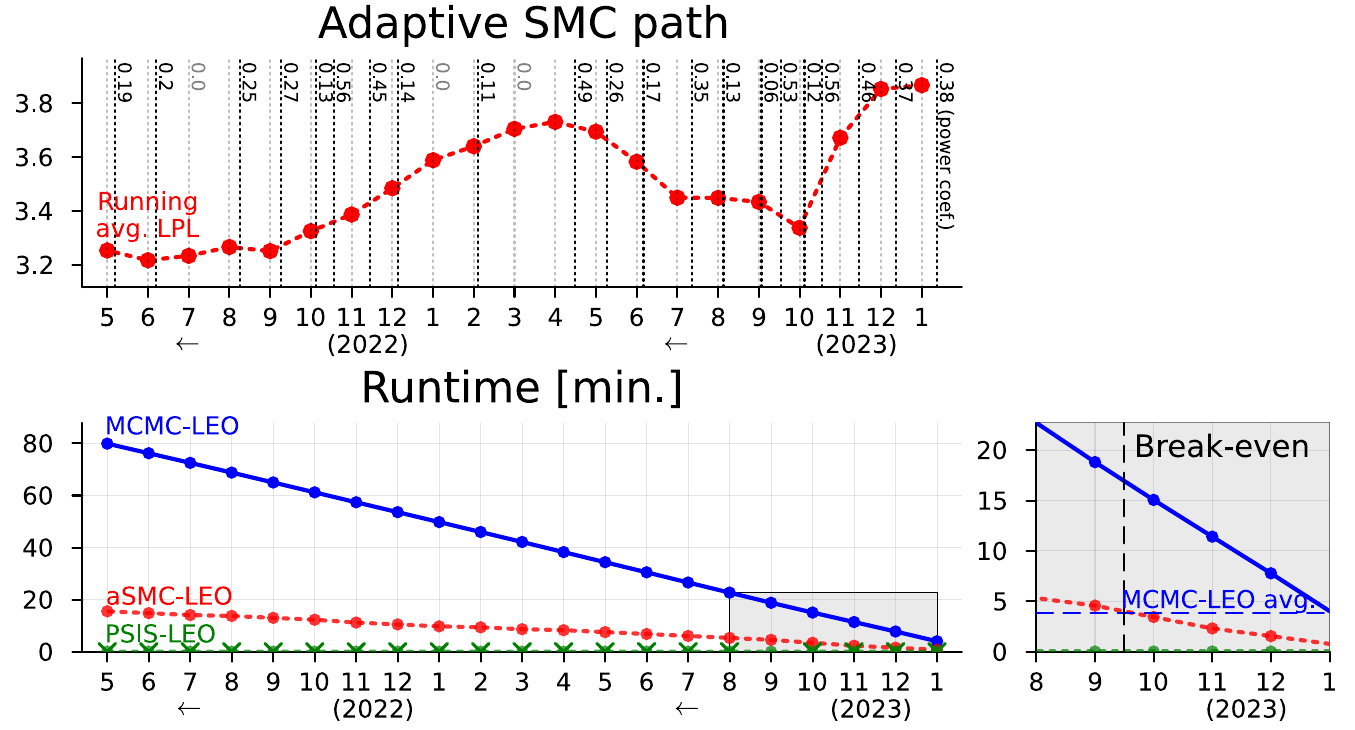}
    \spacingset{1} \caption{
        \textbf{Top}: Intermediate distributions adaptively determined by the aSMC sampler;
        the power coefficient on the corresponding pseudo-latest likelihood contribution is indicated.
        The \textcolor{red}{\textbf{(red) dashed line}} represents the backward running average LPL for aSMC as in Figure \ref{fig:yield-lpd}.
        \textbf{Bottom}: Cumulative run time comparison, interpreted \textit{backwards}, as the latest observations are sequentially deleted backwards.
        For PSIS, we denote the point at which it was diagnosed as $\hat{k} > 0.7$ using a cross ($\times$).
        \textbf{Bottom right}: Zoomed-in.
        The break-even point incates the \textit{furthest} time point from the right at which cumulative aSMC run time remains below the average run time of MCMC.
    }
    \label{fig:yield-paths}
\end{figure}

Figure \ref{fig:yield-paths} shows where the sampler applied the invariant MCMC kernel.
Notably, at some points, no {sub-intermediate MCMC kernel interventions} were required {(e.g., from month 4 to 3 of the year 2023)}, while for others, multiple sub-intermediate interventions were applied {(e.g., from month 11 to 10 of the year 2022)};
the latest data point provides different levels of information over time and the sampler adjusts in response to maintain a sound per-period approximation.

Figure \ref{fig:yield-paths} also presents the cumulative run time for each method.
IS is the fastest, as it involves only re-weighting the samples.
This speed comes at the cost of poor approximation quality, particularly for longer case-deletion horizons, as shown in Figure \ref{fig:yield-lpd}.
To compare MCMC and aSMC, note that per time-point computations in MCMC are parallelizable, whereas aSMC chains the computation due to backward deletion.
The break-even point indicates that aSMC is faster up to four backward-sequential deletions.
Since this exceeds one, the discrepancy in cumulative run time continues to widen.
This further suggests that applying comparable parallelism to aSMC (in complement with MCMC) has the potential to yield faster run times than MCMC alone, while maintaining approximation quality comparable to MCMC.

\subsection{Spatial example} \label{sec:4:m5}

\subsubsection{Panel data of retail goods sales}

We conclude with a spatial modeling example.
The spatial dataset from the M5 competition \citep{MakridakisSpiliotisAssimakopoulos2022} contains unit sales at the item level from ten stores located in California (CA), Texas (TX), or Wisconsin (WI).
Each item is classified within a unique department, which is further classified under a unique product category.
For instance, the item \texttt{HOBBIES\_2\_001} belongs to the department \texttt{HOBBIES\_2} and falls under the category \texttt{HOBBIES}.
An exhaustive list of department identifiers is:
\texttt{HOUSEHOLD\_1},
\texttt{HOUSEHOLD\_2},
\texttt{HOBBIES\_1},
\texttt{HOBBIES\_2},
\texttt{FOOD\_1},
\texttt{FOOD\_2}, and
{\texttt{FOOD\_3}}.
For a complete description of the data, we refer to \href{https://www.kaggle.com/competitions/m5-forecasting-accuracy}{https://www.kaggle.com/competitions/m5-forecasting-accuracy}.

We work with item-level sales trajectories summarized by their average daily change in ten store locations (\texttt{CA\_1} to \texttt{CA\_4}, \texttt{TX\_1} to \texttt{TX\_3}, and \texttt{WI\_1} to \texttt{WI\_3}).
For simplicity, the analysis focuses on items numbered \texttt{001} to \texttt{030} per department, producing a balanced panel of $K = 209$ items across $S = 10$ store locations.
To capture variation in item-level unit sales and their spatial co-movement patterns, we estimate a Bayesian hierarchical model that allows for spatial dependence across observations,
\begin{equation*}
    \bm{y}_{k}
        \ind \mbox{MVN}(\bm{\mu} + \alpha_{g[k]} \bm{1}_S, \bm{\Sigma}), \qquad
    \bm{\mu}
        \sim \mbox{MVN}(\bm{0}, \bm{I}_S), \qquad
    \alpha_g
        \iid \mbox{normal}(0, 1),
\end{equation*}
where
$\bm{y}_k = (y_{k,1}, \ldots, y_{k,S})^\tr$ captures spatial variation in unit sales across $S$ stores for item $k$,
$\bm{\mu}$ denotes store-specific means,
$g[k]$ identifies the department associated with item $k$ (e.g., $k = \texttt{FOOD\_1\_001}$ maps to $g[k] = \texttt{FOOD\_1}$),
$\alpha_{g[k]}$ gives the group (department) mean for item $k$, and
$G=30$ is the total number of departments.
Since the exact store {locations} are not disclosed, the conditional covariance is estimated by $\bm{\Sigma} \sim \textsc{IW}(\nu_0 = 2S, \bm{S}_0 = \bm{I}_S)$.
Under the notation in Section \ref{sec:2}, we may write
$\phi = (\bm{\mu}, \bm{\Sigma})$ and $\theta_g = \alpha_g$,
with each index $k$ corresponding to $(g[k], i[k])$;
each conditionally independent $S$-variate observation $\bm{y}_k$ is identified as $y_{g[k],i[k]}$.

\subsubsection{Group multi-fold cross validation over spatially dependent units}

Several CV designs are possible, and the proposed aSMC framework accommodates each.
\begin{enumerate}[(a)]
    \item \textbf{Multifold}: Given conditional independence over $k$, one may randomly partition item indices and evaluate out-of-sample replication at the unit level.
    \item \textbf{LGO}: Alternatively, one may target department-level generalization via LGO, dropping entire groups (departments), such as the example in Section \ref{sec:4:radon}.
    \item \textbf{Group multifold}: For illustration, we consider a different setting.
    Each item maps to a product department, so we apply a grouped 10-fold CV scheme that ensures coverage of each department across all folds;
    we reframe the generalization exercise from predicting new generic items or items in new departments to predicting new items from \textit{existing} departments across ten spatially dependent store locations.
\end{enumerate}

We compute the predictive likelihood by leaving out the subsets $\mathcal{I}_j$, where $j \in \set{1, \ldots, 10}$.
Each subset $\mathcal{I}_j$ is constructed to contain (approximately) $1/J$ of the item identifiers sampled from each department to ensure a balanced representation.
This leads to the unnormalized weights
\begin{equation*}
    w_j
    = \prod_{k \in \mathcal{I}_j} \mbox{MVN}(\bm{y}_k \cond \bm{\mu} + \alpha_{g[k]} \bm{1}_S, \bm{\Sigma})^{-1}.
\end{equation*}

Given that each subset involves approximately 20 $(S=10)$-variate deletions, a single-step IS approach is unlikely to yield reliable estimates;
the aSMC sampler is therefore applied via the LSO design described in Section \ref{sec:2:2:4} and visualized in Figure \ref{fig:smc-example:lso}.
With the baseline MCMC run taking roughly 12.2 minutes to yield 1000 (thinned and post burn-in) samples, total run time for completing MCMC approximations across all folds can extend up to about 2 hours.
Although not prohibitively expensive, it seems burdensome for evaluating a single model, considering the iterative nature of applied modeling, where diagnostics often inform model extensions or refinements.

\begin{figure}
    \centering\spacingset{1}
    \includegraphics[width=\textwidth]{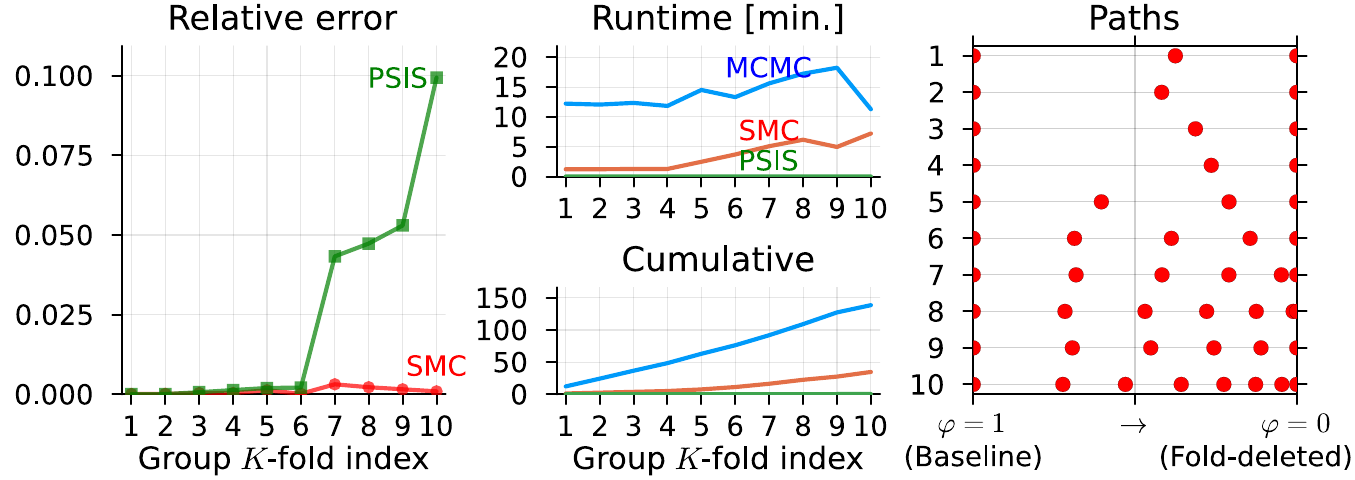}
    \spacingset{1} \caption{
        \textit{Spatial example}.
        Summary of group $K$-fold results.
        \textbf{Left}: Relative error of the log predictive likelihood approximations. Lower relative error indicates that the estimates were closer to the second-best strategy of re-running MCMC to re-approximate the posterior.
        \textbf{Center}: Fold-wise and cumulative comparison of run times.
        \textbf{Right}: The adaptively determined paths of distribution when running aSMC for each of the folds.
    }
    \label{fig:m5}
\end{figure}

Figure \ref{fig:m5} summarizes the results.
We measured the discrepancy between the reference log predictive likelihood, obtained by re-running MCMC for each fold, and the approximations produced by aSMC and PSIS, using relative error.
The aSMC approach yields a consistently lower relative error.
Since all methods allow parallel computation over folds, aSMC offers substantial time savings relative to MCMC while offering near-identical approximation quality.
Some folds in the grouped CV setting appear to have contributed to larger posterior shifts, with relative errors occasionally reaching 5--10\% under PSIS.
For these folds, the automatically constructed intermediate distributions in aSMC helped maintain low error.

\section{Summary and discussion} \label{sec:5}

We have introduced an aSMC sampler to (cross) validate Bayesian hierarchical models.
The method was motivated by a computational challenge in Bayesian hierarchical setups, particularly when case-deletion schemes are applied to one or more groups (or subsets) involving multiple and/or non-factorizable observations.
In such scenarios, conventional IS-based approximations can be unreliable, as only an inadequate fraction of the finite posterior draws from the non-case-deleted distribution lie in the higher-mass regions of the case-deleted posterior.
Re-running MCMC becomes the second-best fallback option, which itself is often costly and impracticable with modern complex Bayesian hierarchical models.

The algorithm was designed to be automatic in the sense that
(a) the user need not specify the path of distributions as input,
(b) the selection of (PS)IS and MCMC re-runs is automatically determined,
and
(c) MCMC re-runs targeting the adaptively chosen path bridging the baseline and case-deleted posterior are implemented as an efficient, parallelizable component of the algorithm.
Using three real data examples involving
LGO CV,
group $K$-fold CV, and
LEO validation,
we have shown that the sampler can efficiently and automatically approximate diverse CV schemes and
facilitate the Bayesian workflow.

Although the sampler was designed to be user-friendly, several parameters must still be specified in advance.
For example, the number of times the MCMC kernel is applied currently needs to be set manually by the user.
This presents a trade-off; more iterations may yield higher-quality samples due to the asymptotic exactness of invariant kernels, but they also increase run time.
We have advocated
leveraging parallel resources where available and
repurposing information from baseline posterior draws for efficient tuning at no additional cost
(e.g., using autocorrelations to determine the minimal sufficient number of MCMC iterations, and
covariances to initialize the HMC mass matrix).
Whether this strategy is optimal remains unclear, and alternative approaches are worth exploring.
\cite{MargossianHoffmanSountsovRiou-DurandVehtariGelman2024} propose diagnostics for parallel MCMC convergence in a high-chain, low-iteration regime, which is structurally analogous to the SMC setup with many particles and parallelized updates.
Incorporating such diagnostics to develop a tuning-free, user-friendly and more trustworthy algorithm is a promising direction for future refinement.

\newpage

\if0\blind
{

\section*{Acknowledgments}
The majority of this research was carried out while GH was a graduate student at Columbia University GSAS.
AG's work was partially supported by the Office of Naval Research grant number N000142212648.

\section*{Disclosure Statement}
There are no competing interests to declare.

} \fi

\if1\blind
{
} \fi

\bigskip
\begin{center}
{\large\bf SUPPLEMENTARY MATERIAL}
\end{center}

\begin{description}

\item[\texttt{Appendix.pdf}:] Brief expositions of
(a) Gibbs sampler (\dquote{Gibbs sampler for the dynamic Nelson--Siegel (DNS) model}) and
(b) sensitivity analysis (\dquote{Additional sensitivity analysis}).
(PDF file)

\item[\texttt{DynamicImage.gif}:] A dynamically rendered image  accompanying the sensitivity analysis in the appendix (subsection \dquote{Additional sensitivity analysis}).
(GIF file)

\item[\texttt{aSMC-BayesCV}:] The folder includes files (data, Julia code, and outputs) required to run the methodology and reproduce results presented in the article.
Please refer to \texttt{README.md} (\href{https://github.com/geonhee619/ASMC-BayesCV}{https://github.com/geonhee619/ASMC-BayesCV}) for complete descriptions of setup instruction and execution flow.
(folder)

\end{description}

\break

\bibliographystyle{apalike}
\bibliography{ref.bib}

\end{document}


\def\spacingset#1{\renewcommand{\baselinestretch}%
{#1}\small\normalsize} \spacingset{1}

\if0\blind
{
  \title{\bf Appendix for: \\ Adaptive sequential Monte Carlo for structured cross validation in Bayesian hierarchical models}
  \author{
    Geonhee Han\thanks{gh2610@columbia.edu.}
    \hspace{.2cm} \\
    \small Graduate School of Arts and Sciences, Columbia University\thanks{The majority of this research was carried out while GH was a graduate student at Columbia University GSAS QMSS.} \\
    \small Graduate School of Public Policy, The University of Tokyo \\
    \\
    Andrew Gelman \hspace{.2cm} \\
    \small Department of Statistics and Department of Political Science, Columbia University
    }
    \date{11 Aug 2025}
    \maketitle
} \fi

\if1\blind
{
  \bigskip
  \bigskip
  \bigskip
  \begin{center}
    {\LARGE\bf Title}
\end{center}
  \medskip
} \fi

\spacingset{1.1}
\renewcommand{\arraystretch}{0.6}


\appendix
\section{Supplementary} \label{sec:appendix}

\subsection{Gibbs sampler for the dynamic Nelson--Siegel (DNS) model}

The DNS model considered in the manuscript admits a representation as a normal dynamic linear model:
\begin{align*}
    \bm{y}_t &= \bm{X}_t \bm{\beta}_t + \bm{\varepsilon}_t^{(y)}, \qquad
    \bm{\varepsilon}_t^{(y)} \ind \mbox{MVN}(\bm{0}, \bm{\Sigma}^{(y)}), \\
    \bm{\beta}_t &= \bm{\beta}_{t-1} + \bm{\varepsilon}_t^{(\beta)}, \qquad
    \bm{\varepsilon}_t^{(\beta)} \ind \mbox{MVN}(\bm{0}, \bm{\Sigma}^{(\beta)}),
\end{align*}
where $\bm{\beta}_0 \sim \mbox{MVN}(\bm{m}, \bm{P}^{-1})$,
$\bm{\Sigma}^{(y)} \sim \textsc{IW}(\nu_0^{(y)}, \bm{S}_0^{(y)})$, and
$\bm{\Sigma}^{(\beta)} \sim \textsc{IW}(\nu_0^{(\beta)}, \bm{S}_0^{(\beta)})$.
The design matrix $\bm{X}_t = [\bm{x}_{t}(\tau_1), ..., \bm{x}_{t}(\tau_K)]^\tr$ is defined via
\begin{equation*}
    \bm{x}_{t}(\tau) = \begin{bmatrix}
        1 \\
        {1 - \exp(-\lambda_t \tau) \over \lambda_t \tau} \\
        {{1 - \exp(-\lambda_t \tau) \over \lambda_t \tau} - \exp(-\lambda_t \tau)}
    \end{bmatrix}.
\end{equation*}
Suppose the index runs over $t = 1,\dots,T$.
The joint density is
\begin{equation*}
    p(\bm{\Sigma}^{(y)}) p(\bm{\Sigma}^{(\beta)}) p(\bm{\beta}_0) \prod_{t=1}^T p(\bm{\beta}_t \cond \bm{\beta}_{t-1}, \bm{\Sigma}^{(\beta)}) p(\bm{y}_t \cond \bm{\beta}_t, \bm{\Sigma}^{(y)}).
\end{equation*}
In order to implement continuous case deletions, it suffices to consider sampling from the posterior distribution proportional to the joint density with the likelihood contribution at $T$ controlled by a power scaling coefficient $\rho \in [0,1]$:
\begin{equation*}
    p(\bm{\Sigma}^{(y)}) p(\bm{\Sigma}^{(\beta)}) p(\bm{\beta}_0) \bracket{\prod_{t=1}^T p(\bm{\beta}_t \cond \bm{\beta}_{t-1}, \bm{\Sigma}^{(\beta)})}
    \bracket{\prod_{t=1}^{T-1} p(\bm{y}_t \cond \bm{\beta}_t, \bm{\Sigma}^{(y)})}
    p(\bm{y}_T \cond \bm{\beta}_T, \bm{\Sigma}^{(y)})^{\rho}
    .
\end{equation*}
If $\rho = 1$, this reduces to a normal dynamic linear model where the index runs over $t = 1,\dots,T$.
If $\rho = 0$, the same is true where the index now runs over $t = 1,\dots,T-1$.
Hence, let $\rho \in (0,1)$.
Note
\begin{equation*}
    p(\bm{y}_T \cond \bm{\beta}_T, \bm{\Sigma}^{(y)})^{\rho}
    \propto
    |\bm{\Sigma}^{(y)}|^{-\rho/2}
    \exp\!\left(
        -{\rho \over 2}
        (\bm{y}_T - \bm{X}_T\bm{\beta}_T)^\tr
        [\bm{\Sigma}^{(y)}]^{-1}
        (\bm{y}_T - \bm{X}_T\bm{\beta}_T)
    \right).
\end{equation*}
The Gibbs sampler iterates the following steps.
We use the notation $-$ to abbreviate conditioning upon all other variables.
\begin{enumerate}
    \item Draw $(\bm{\beta}_0 \cond -) \sim \mbox{MVN}(\hat{\bm{m}}, \hat{\bm{P}}^{-1})$, where
    \begin{align*}
        \hat{\bm{P}} &= \bm{P} + (\bm{\Sigma}^{(\beta)})^{-1}, \\
        \hat{\bm{m}} &= \hat{\bm{P}} (\bm{P}\bm{m} + (\bm{\Sigma}^{(\beta)})^{-1} \bm{\beta}_1).
    \end{align*}

    \item Draw $(\bm{\Sigma}^{(\beta)} \cond -) \sim \textsc{IW}(\nu_0^{(\beta)} + T, \bm{S}_0^{(\beta)} + \sum_{t=1}^{T} (\bm{\beta}_t - \bm{\beta}_{t-1}) (\bm{\beta}_t - \bm{\beta}_{t-1})^\tr)$.

    \item Draw $(\bm{\beta}_{1:T} \cond -)$ using the forward-filtering backward-sampling algorithm \citep{CarterKohn1994, FruhwirthSchnatter1994}, but modifying the measurement covariance as $\rho^{-1} \bm{\Sigma}^{(y)}$ at $t = T$.

    \item Draw $(\bm{\Sigma}^{(y)} \cond -) \sim \textsc{IW}(\hat{\nu}, \hat{\bm{S}})$, where
    \begin{align*}
        \hat{\nu} &= \nu_0^{(y)} + T - (1 - \rho), \\
        \hat{\bm{S}} &= \bm{S}_0^{(y)} + \sum_{t=1}^{T-1} (\bm{y}_t - \bm{X}_t\bm{\beta}_t) (\bm{y}_t - \bm{X}_t\bm{\beta}_t)^\tr + \rho (\bm{y}_T - \bm{X}_T\bm{\beta}_T) (\bm{y}_T - \bm{X}_T\bm{\beta}_T)^\tr.
    \end{align*}
\end{enumerate}

\newpage

\subsection{Additional sensitivity analysis}

Traditional workflows and recommended practices \citep{Millar2018, BurknerGabryVehtari2020, VehtariSimpsonGelmanYaoGabry2024} often resort to re-running MCMC when the distribution of importance weights via the generalized Pareto statistic $\hat{k} > 0.7$ suggests unreliable estimates.
The aSMC sampler output in this article is similarly conditioned upon $\hat{k} > 0.7$.
The rationale for this is to
(1) design the algorithm as a proper generalization of the original PSIS-LOO-CV \citep{VehtariGelmanGabry2017}, and
(2) appropriately assess the reliability of the estimates.
As stated in the main text, the difference is that the re-run is integrated as a design-efficient component of the algorithm, in the sense that they reduce to parallelizable application(s) of an invariant MCMC kernel.

It would be useful to briefly assess the sensitivity of the algorithm with respect to the choice of threshold $0.7$.
The accompanying GIF (also at \href{https://github.com/geonhee619/ASMC-BayesCV/blob/main/img/DynamicImage.gif}{https://github.com/geonhee619/ASMC-BayesCV/blob/main/img/DynamicImage.gif}) dynamically visualizes the results from the LSO-CV example using five thresholds:
\begin{itemize}
    \item $-\infty$ (always trigger invariant kernel intervention),
    \item $0.5$ (more stringent),
    \item $0.7$ (default),
    \item $0.9$ (more lenient), and
    \item $+\infty$ (never trigger).
\end{itemize}
The random seed is kept constant throughout to isolate the effect of triggering kernel interventions.

Empirically, the procedure appears to be robust to the choice of threshold, and the graphically supported interpretation is as follows.
The diagnostic $\hat{k}$ is applied exclusively in the \textit{final step}, where finally the sampler seeks to draw from the fold-deleted posterior from the previous draws.
At this point, again, if $\hat{k} < 0.7$, we proceed with PSIS instead of MCMC, with the rationale that the preceding steps may already have yielded samples from a distribution sufficiently close to the fold-deleted target (e.g., tempering coefficient of $0.01$).
Because the threshold affects only this final decision point, more lenient values may indeed marginally increase the estimation error.
However, the degradation is negligible relative to simple importance weighting, as previous partially case-deleted draws (often with much greater overlap to the target) are already available.

Now, a natural question is whether this is necessary at all if the results are insensitive.
Despite the insensitivity, incorporating this diagnostic step
(a) may offer run time gains when simple importance weighting suffices, and
(b) is, again, important to appropriately assess the reliability of the estimates otherwise.

\break

\bibliographystyle{apalike}
\bibliography{ref.bib}